\documentclass[
reprint,
superscriptaddress,
nofootinbib,
amsmath,amssymb,
aps,
prstab,
floatfix,
]{revtex4-2}

\usepackage{graphicx}
\usepackage{dcolumn}
\usepackage{bm}
\usepackage[T1]{fontenc}
\usepackage{newtxmath,newtxtext}
\usepackage{mathtools}
\usepackage{physics}
\usepackage{enumitem}
\usepackage[caption=false]{subfig}
\usepackage{derivative}
\usepackage{natbib}
\usepackage{xr-hyper}
\usepackage{hyperref}
\usepackage{cancel}
\usepackage{xcolor}
\usepackage{makecell}
\usepackage{float}
\usepackage{svg}
\usepackage[para,online,flushleft,referable]{threeparttablex}

\usepackage{ragged2e} 
\DeclareCaptionJustification{justified}{\justifying}

\makeatletter
\def\bibsection{%
  \par
  \begingroup
    \baselineskip26\p@
    \bib@device{\hsize}{72\p@}%
  \endgroup
  \nobreak\@nobreaktrue
  \addvspace{19\p@}%
  }%
\makeatother

\begin{document}
\preprint{APS/123-QED}
\title{Impact of beam asymmetries at the Future Circular Collider $e^+e^-$}

\author{Peter Kicsiny}
\email{peter.kicsiny@cern.ch}
\affiliation{European Organisation for Nuclear Research (CERN), CH 1211 Geneva 23, Switzerland}
\affiliation{\'Ecole Polytechnique F\'ed\'erale de Lausanne (EPFL), Route Cantonale, 1015 Lausanne, Switzerland}

\author{Xavier Buffat}
\affiliation{European Organisation for Nuclear Research (CERN), CH 1211 Geneva 23, Switzerland}

\author{Khoi Le Nguyen Nguyen}
\affiliation{Cavendish Laboratory, University of Cambridge, JJ Thomson Avenue, Cambridge CB3 0HE, UK}

\author{Tatiana Pieloni}
\affiliation{\'Ecole Polytechnique F\'ed\'erale de Lausanne (EPFL), Route Cantonale, 1015 Lausanne, Switzerland}

\author{Mike Seidel}
\affiliation{\'Ecole Polytechnique F\'ed\'erale de Lausanne (EPFL), Route Cantonale, 1015 Lausanne, Switzerland}
\affiliation{Paul Scherrer Institut (PSI), Forschungsstrasse 111, 5232 Villigen, Switzerland}

\date{\today}

\begin{abstract}
In this paper we present detailed simulations with asymmetric initial beam settings in the context of the proposed Future Circular Collider $e^+e^-$ (FCC-ee) using the \texttt{Xsuite} framework. We compare simulated equilibrium bunch sizes and luminosities against an already existing analytical model, which shows remarkably good agreement for realistic small perturbations. We investigate the longitudinal top-up injection, the currently preferred injection scheme for the FCC-ee, using self-consistent simulations featuring beam-beam collisions with beamstrahlung and the injection process, for the first time. We present and assess the sensitivity and required precision of the nominal beam parameters in a potential real-life operation by providing first estimates of the tolerances in the initial asymmetry of several machine parameters, with respect to the 3D flip-flop mechanism, obtained from parameter scan simulations.
\end{abstract}

\maketitle


\section{\label{sec:1}Introduction}

The Future Circular Collider (FCC)~\cite{Benedikt:2651299} is currently the most preferred next generation particle collider project at CERN. Its first stage consists of an electron-positron collider, called the FCC-ee. For this machine, the beam lifetime will be determined by the dynamic aperture and by radiation in the machine arcs and during collisions. In the latter case, radiation can be produced by scattering in a single particle's electromagnetic field, which is called radiative Bhabha scattering~\cite{Bhabha:1936zz}. The corresponding energy loss of the beam particles leads to a continuously decreasing beam intensity and luminosity. Contrary to this incoherent, single-particle scattering, in this paper we focus on photon emission due to the bending in the collective electromagnetic field of the opposing beam, referred to as beamstrahlung~\cite{Yokoya:1991qz}. As compared to radiative Bhabha scattering, here the energy loss of the primary particle is typically smaller and the deflected primaries mostly stay within the machine aperture. Nevertheless, the beam properties at equilibrium, and in particular the bunch length and the energy spread, will be significantly altered by beamstrahlung~\cite{Shatilov:2806204, PhysRevLett.110.114801}. Since beamstrahlung depends on the magnitude of the beam-beam force, and thus on the properties of the two colliding beams, the equilibrium emittances are no longer defined by the static properties of the lattice but are rather dominated by beam properties which are potentially dynamic.

Due to the low beam lifetime of the FCC-ee, the collider is designed to operate in top-up injection mode~\cite{Aiba:2815853} using a single booster ring~\cite{Ogur:2301994}, whereby low intensity bunches are injected with a frequency of $\sim$0.1~Hz~\cite{zimmermann2014fcc} to the stored higher intensity bunch, in order to compensate for the decaying bunch population. Since there is only one booster ring to feed the two collider rings, only one of the two beams can be topped-up at a given time, thus an asymmetry between the intensity of the two beams always exists. In general, the longer the time between two consecutive injections, the higher the asymmetry. It is therefore important to determine the tolerance on the asymmetry between the two beams from the beam dynamics point of view as it may potentially constrain the specification for the booster repetition rate.

It was observed in simulations~\cite{Shatilov:2816655} that such an intensity asymmetry can trigger a mechanism, called 3D flip-flop, in which the size of the initially lower intensity bunch blows up, while that of the other one shrinks to the parameters defined by the machine lattice. The effect is caused by the combination of beamstrahlung and the strong coupling between the transverse and the longitudinal planes caused by the large crossing angle between the two beams at collisions. Indeed, the transverse beam-beam force and consequently the beamstrahlung will be affected by changes in the bunch lengths while the strength of beamstrahlung determines the bunch length, possibly leading to a runaway situation.

In general, the flip-flop mechanism refers to a dynamical behavior resulting in one of two possible equilibrium states, in which one of the colliding bunches blows up in size while the other bunch shrinks or remains at the same size. Purely transverse flip-flop has already been studied in the '70s, at the SPEAR~\cite{donald1979investigation, Richter:1969wy, Tennyson}, and at the PEP~II~\cite{holtzapple2002observation} $e^+e^-$ collider rings. The naming of the effect originates from the fact that in that case it was possible to flip the beam system out of one stable asymmetric equilibrium into the other by varying the radiofrequency phasing in the cavities on either side of the interaction point (IP)~\cite{Tennyson}. A purely transverse flip-flop effect has also been observed in the past at the VEPP-2000 collider~\cite{shwartz2014recent}. In these cases, the mechanism was triggered by nonlinear behavior and radiation did not play a role. By contrast, in case of the FCC-ee, the flip-flop effect can be triggered due to beamstrahlung, by an initial small asymmetry in the beam or machine parameters.

Let us consider a bunch with a lower intensity than its opposing bunch at a collision, such that the opposing bunch undergoes weaker beamstrahlung, resulting in the shortening of that bunch, with respect to the equilibrium bunch length with symmetric intensities. Due to this shortening, the low intensity bunch will now undergo stronger beamstrahlung, as the intensity of the other bunch is now confined into a shorter length. This results in an increase of the length of the low intensity bunch. This process may self-enhance as the low intensity bunch lengthens and the high intensity bunch shortens, thus further weakening and strengthening beamstrahlung respectively. For sufficiently small asymmetries, the two beams will reach a new equilibrium with asymmetric bunch intensities and lengths. On the other hand, for sufficiently large asymmetries, the mechanism saturates as the high intensity bunch reaches the equilibrium given by the lattice properties, corresponding to a configuration of negligible beamstrahlung for that bunch.

Before reaching this saturation, an additional unstable mechanism kicks in, usually above a certain threshold: the increase of the strength of the beam-beam force experienced by the low intensity bunch enables a strong nonlinear diffusion mechanism leading to its blowup in the transverse directions as well as particle losses. This further enhances the instability mechanism by reducing beamstrahlung for the high intensity bunch. Transverse beam size asymmetries could lead to a similar behavior as intensity asymmetries, by increasing and reducing beamstrahlung for the larger and smaller beam respectively. Given that the vertical emittance of each beam is solely defined by the quality of the optics correction in the two collider rings, the tolerance on the vertical beam size asymmetry may constrain the specification for optics tuning.

The contributions of this paper are a set of numerical studies in the context of the FCC-ee design. On one hand, we present a set of benchmark studies of simulated equilibrium bunch lengths and luminosities, under asymmetric bunch intensities, against estimates with an analytical model, which was developed in~\cite{nguyen2024impact}. On the other hand, we present various parameter scans of the beam and machine parameters, and make first estimates for tolerances in the asymmetry of these parameters with respect to the 3D flip-flop instability. The paper is organized as follows: it starts with a presentation of the analytical model in Sec.~\ref{sec:2}, which can estimate the equilibrium bunch length for asymmetric configurations. In Sec.~\ref{sec:3}, we detail our numerical model. In Sec.~\ref{sec:4}, we present our estimates for bunch lengths and luminosity, obtained from tracking, and compare them against the analytical predictions. In Sec.~\ref{sec:5}, we perform simulations of the longitudinal top-up injection process, and discuss the beam dynamics under different types of perturbations in the beam parameters. Finally in Sec.~\ref{sec:6} we summarize our findings.

\section{\label{sec:2}Analytical estimation of equilibrium bunch length}

An analytical formalism to estimate the equilibrium bunch length in configurations featuring asymmetric beam properties is not a contribution of this paper. It has been developed in~\cite{nguyen2024impact} and it follows the approach developed in~\cite{Garcia} for symmetric configurations. The main results are summarized in this section and used in the later sections for benchmarking simulation results.

Throughout this paper the subscript $l$ denotes the parameters of the low intensity bunch while the subscript $h$ denotes those of the high intensity bunch. In this section only, all coordinates $\{x,y,z\}$ and r.m.s. bunch sizes $\{\sigma_{x}^*, \sigma_y^*, \sigma_z\}$ are understood in a Lorentz transformed (boosted) reference system, where the collision is head-on~\cite{PhysRevLett.74.2228}. The starting point of the derivation is the time dependent bending radius of a single particle of the low intensity bunch $\rho_l(s,t)$ traversing the electromagnetic field of the high intensity bunch. The expression for $\rho_l(s,t)$ can be simplified by making two approximations. The first approximation is to take the flat beam limit ($\sigma_{x,h}^* \gg \sigma_{y,h}^*$) which applies to the FCC-ee design parameters. The second approximation is to restrict the treatment to small amplitude particles, i.e. $|x|<\sigma^*_{x,h}$ and $|y|<\sigma^*_{y,h}$. Furthermore, we neglect the impact of the hourglass effect~\cite{Herr:941318} and crab-waist optics~\cite{Zobov_2016}. With these assumptions, the bending radius can be written as:
\begin{align}
	&\frac{1}{\rho(x,y,z,s)_l}\approx\sqrt{\frac{2}{\pi}}\frac{2r_eN_h}{\gamma_l\sigma_{z,h}}\nonumber \\&\times\frac{\exp[-x^2/(2\sigma_{x,h}^{*2})]}{\sigma_{x,h}^*}\exp\left[-\frac{(2s-z)^2}{2\sigma_{z,h}^2}\right]\nonumber\\
	&\times\left[\left(\frac{y}{\sigma_{y,h}^*}\right)^2+\left(\frac{x-(s-z/2)\theta_c}{\sigma_{x,h}^*}\right)^2\right]^{1/2},
\label{eq:1}
\end{align}
where $\theta_c$ denotes the full crossing angle at collision, $r_e$~$\approx$~$2.818\cdot 10^{-15}$~[m] the classical electron radius, $\gamma_l$ the relativistic Lorentz factor of the low intensity bunch and $N_h$ the number of elementary charges in the high intensity bunch. Integrating powers of the local curvature over the longitudinal coordinate $s$ we obtain:
\begin{align}
	&\mathcal{I}_{n,l}\equiv\int _{-\infty}^{+\infty}ds\Bigg\langle\dfrac{1}{\rho_l^n}(s)\Bigg\rangle\nonumber\\
	&=\dfrac{1}{(2\pi)^{3/2}}\iiiint\limits_{x,y,z,s\in\mathbb{R}^4}\dfrac{dxdydzds}{\sigma_{x,l}^*\sigma_{y,l}^*\sigma_{z,l}}\rho(x,y,z,s)_l^{-n}\nonumber\\
	&\times\exp\left[-\dfrac{\left(x+{z\theta_c}/{2}\right)^2}{2\sigma_{x,l}^{*2}}-\dfrac{y^2}{2\sigma_{y,l}^{*2}}-\dfrac{z^2}{2\sigma_{z,l}^2}\right].
\label{eq:2}
\end{align}
The solution for $\mathcal{I}_{n,l}$ in Eq.~\eqref{eq:2} can be expressed with hypergeometric integrals~\cite{Wang&Guo}, which can be solved numerically.

The evolution of the longitudinal bunch size $\sigma_{z,l}$ of the low intensity bunch over time can be described by a differential equation featuring the growth caused by quantum excitation as well as the radiation damping each arising from synchrotron radiation in the lattice and beamstrahlung:
\begin{align}
	\odv{\sigma_{z,l}^2}{t}&=\dfrac{2}{\tau_{z,\text{SR},l}}(\sigma_{z,\text{SR},l}^2+\mathcal{A}_{l}\mathcal{I}_{3,l})\nonumber\\
	&\hspace{10pt}-\left(\dfrac{2}{\tau_{z,\text{SR},l}}+\mathcal{B}_{l}\mathcal{I}_{2,l}\right)\sigma_{z,l}^2,
\label{eq:6}	
\end{align}
with $\tau_{z,\text{SR/BS},l}$ being the longitudinal damping times of the r.m.s. bunch length coming from synchrotron radiation and beamstrahlung, respectively. The beamstrahlung damping time is defined as:
\begin{equation}
    \tau_{z,\text{BS},l} =\frac{E_l}{U_{\text{BS},l}}.
    \label{eq:tau_sr_bs}
\end{equation}
The energy loss due to beamstrahlung $U_{\text{BS},l}$ can be defined as:
\begin{equation}
    U_{\text{BS},l}=\mathcal{N}_{\text{ph},l} \delta_{\text{BS},l} E_l.
    \label{eq:u_bs_asym}
\end{equation}
In Eq.~\eqref{eq:u_bs_asym} we used the average number of emitted beamstrahlung photons in a collision, approximated as:
\begin{equation}
\label{eq:n_avg_asym}
\mathcal{N}_{\text{ph},l}\approx\dfrac{5}{2\sqrt{3}}\alpha\gamma_l\mathcal{I}_{1,l},
\end{equation}
with $\alpha$ being the fine structure constant. Furthermore, the average relative energy loss in a single collision by a single primary can be approximated as:
\begin{equation}
\label{eq:delta_avg_asym}
\delta_{\text{BS},l}\approx\dfrac{2}{3}r_e\gamma_{l}^3\mathcal{I}_{2,l}.
\end{equation}
The constants $\mathcal{A}_{l}$ and $\mathcal{B}_{l}$ are expressed as:
\begin{equation}	
\mathcal{A}_{l}\equiv\frac{n_\text{IP}\tau_{z,\text{SR},l}}{4T_\text{rev}}\left(\frac{\alpha_p C}{2\pi Q_\text{s}}\right)^2\frac{55}{24\sqrt{3}}\frac{r_e^2\gamma_{l}^5}{\alpha},
\end{equation}
\begin{equation}
\mathcal{B}_{l}\equiv n_\text{IP}\frac{4}{3}r_e\gamma_{l}^3.
\end{equation}
Here $n_{\text{IP}}$ denotes the number of IPs in the collider ring, $T_{\text{rev}}$ the revolution time, $\alpha_p$ the momentum compaction factor, $C$ the collider circumference and $Q_s$ the synchrotron tune. These constants express the dependence of the dynamics on the machine and beam parameters. By writing up Eq.~\eqref{eq:6} for both interacting bunches one obtains a system of two equations which are coupled through the terms $\mathcal{I}_{n,l}$. This system can be solved iteratively for the equilibrium bunch lengths, when $\odv{\sigma_{z,l/h}^2}{t}=0$ with the subscripts referring to the low or high intensity bunch, respectively.

\section{\label{sec:3}Tracking simulations}

\subsection{\label{sec:31}Beam-beam model}

The simulations have been performed with the \texttt{Xsuite} framework~\cite{iadarola2023xsuite, xsuite}, which is a general purpose multiparticle tracking tool. This section presents the modeling of beam-beam collisions briefly. A detailed description can be found in~\cite{moplo063}. 

In the rest of this paper, all r.m.s. bunch sizes $\{\sigma_{x}^*, \sigma_y^*, \sigma_z\}$ are understood in the unboosted accelerator frame. To account for the effect of boost in the following formulae, the effective horizontal r.m.s. size at the IP $\sigma_{x,\text{eff}}^*$ and the interaction length $L_i$ are introduced as:
\begin{equation}
    \sigma_{x,\text{eff}}^* = \sigma_x^*\sqrt{1 + \Phi^2},
\end{equation}
and
\begin{equation}
    L_i = \frac{\sigma_{z,\text{BS}}}{\sqrt{1 + \Phi^2}}.
\label{eq:li}
\end{equation}
The quantity $\Phi$ is the Piwinski angle, given by
\begin{equation}
\label{eq:piwi}
    \Phi=\frac{\sigma_{z,\text{BS}}}{\sigma_x^*}\tan\left(\frac{\theta_c}{2}\right).
\end{equation}
The beam-beam model is based on an approach developed in~\cite{PhysRevLett.74.2228}, in which a collision with a crossing angle is treated by performing a rotation and a Lorentz boost into a head-on frame, thus simplifying the mathematical description of the electromagnetic field of the bunches. The bunches are then sliced longitudinally and moved across each other in discrete steps. At each step the force represented by the single slice is computed using the soft-Gaussian approximation, meaning that the force is calculated for a Gaussian distribution~\cite{Bassetti:122227} based on the statistical properties of the particles in the slice. In this study, the statistical properties of the slices are evaluated periodically, for all slices at the beginning of each 100th collision (quasi-strong-strong approach). This reduces the computational load with respect to a full strong-strong case, in which these moments are recomputed at each collision and after each slice pair interaction. This is well justified by the fact that the beam parameters are slowly varying over several turns in the instability mechanism considered as well as by the low disruption parameters $D_{x,y}$ of the bunch, defined as:
\begin{equation}
    D_{x} = \frac{2Nr_e}{\gamma}\frac{L_i}{\sigma_{x,\text{eff}}^*(\sigma_{x,\text{eff}}^* + \sigma_{y}^*)},
    \label{eq:dx}
\end{equation}
\begin{equation}
    D_{y} = \frac{2Nr_e}{\gamma}\frac{L_i}{\sigma_{y}^*(\sigma_{x,\text{eff}}^* + \sigma_{y}^*)}.
    \label{eq:dy}
\end{equation}
These dimensionless parameters describe the inverse focal length of the transverse trajectory in units of the bunch length~\cite{chen2005introduction, HOLLEBEEK1981333}. They give an estimate for the number of betatron oscillations performed by a single particle during the collision, with a small value indicating that the beam properties are not significantly distorted during the interaction. For the FCC-ee, the disruption parameters are close to or below 1, hence justifying the usage of the quasi-strong-strong model.

The parameters used in our simulations presented here are summarized in Tab.~\ref{tab:FCC}. Some parameters have been calculated using equations presented in this paper, while the others are taken from~\cite{oide_fccweek}. Based on the dynamic aperture obtained from nonlinear tracking, the t\=t momentum acceptance is asymmetric: $^{+2.5}_{-2.8}~\%$~\cite{oide_fccweek}. As a conservative simplification, we used $\pm2.5$~\% in our linear tracking simulations.

\begin{table}[h!]
\caption{\label{tab:FCC}Selected parameters of the FCC-ee 4 IP baseline design, taken from~\cite{oide_fccweek}, otherwise indicated.}
\begin{threeparttable}
    \begin{tabular}{lcccc}
\hline\hline
  & \textbf{Z} & \textbf{W$^{\pm}$}& \textbf{ZH} & \textbf{t\=t}\\
    \hline
                        $C$ [km] & \multicolumn{4}{c}{90.658816} \\
               $\theta_c$ [mrad] & \multicolumn{4}{c}{30} \\
                       $E$ [GeV] &  45.6 &    80 &   120 & 182.5 \\
               $N_0$ [$10^{11}$] &  1.51 &  1.45 &  1.15 &  1.55 \\
          $\alpha_p$ [$10^{-6}$] &  28.6 &  28.6 &   7.4 &   7.4 \\
                 $\beta_x^*$ [m] &  0.11 &  0.22 &  0.24 &     1 \\
                $\beta_y^*$ [mm] &   0.7 &     1 &     1 &   1.6 \\
         $\sigma_{x}^*$ [$\mu$m] &  8.84 & 21.85 & 13.05 & 39.87 \\
             $\sigma_{y}^*$ [nm] & 22.91 & 35.35 & 29.15 & 37.95 \\
     $\sigma_{z,\text{SR}}$ [mm] &   5.6 &  3.47 &   3.4 &  1.81 \\
     $\sigma_{z,\text{BS}}$ [mm] &  12.7 &  5.41 &   4.7 &  2.17 \\
    $\sigma_{\delta,\text{SR}}$ [$10^{-4}$] &  3.9 &     7 & 10.4 &   16 \\
    $\sigma_{\delta,\text{BS}}$ [$10^{-4}$] &  8.9 &  10.9 & 14.3 & 19.2 \\
        $\Phi$ [1]\tnotex{1} &   21.56 &    3.71 &    5.40 &    0.82 \\
         $Q_x$ [1] & 218.158 & 218.186 & 398.192 & 398.148 \\
         $Q_y$ [1] &   222.2 &  222.22 & 398.358 & 398.182 \\
         $Q_s$ [1] &   0.029 &   0.081 &   0.032 &   0.091 \\
         Momentum acceptance [\%] &     $\pm$ 1 &      $\pm$ 1 &    $\pm$ 1.6 &    $\pm$ 2.5 \\
    $U_{\text{SR}}$ [GeV] &   0.039 &   0.037 &    1.89 &   10.42 \\
    $U_{\text{BS}}$ [MeV]\tnotex{2} &     0.5 &    3.06 &    9.84 &   38.27 \\
    $\tau_{z,\text{SR}}$ [turns]\tnotex{3} &  1157 &   214 &    63 &   18 \\
    $\tau_{z,\text{BS}}$ [turns]\tnotex{3} & 90223 & 26074 & 12190 & 4769 \\
    $D_x$ [$10^{-3}$]\tnotex{4} & 0.15 & 1.04 & 0.46 & 1.55 \\
            $D_y$ [1]\tnotex{5} & 0.94 & 1.86 & 0.88 & 1.58 \\
    $\xi_x$ [1] & 0.0023 & 0.013 & 0.010 & 0.073 \\
    $\xi_y$ [1] &  0.096 & 0.128 & 0.088 & 0.134 \\
\hline\hline
\end{tabular}
\begin{tablenotes}
     \item[1] \label{1} Eq.~\eqref{eq:piwi}.
     \item[2] \label{2} Eq.~\eqref{eq:u_bs_asym}.
     \item[3] \label{3} Eq.~\eqref{eq:tau_sr_bs}.
     \item[4] \label{4} Eq.~\eqref{eq:dx}.
     \item[5] \label{5} Eq.~\eqref{eq:dy}.
\end{tablenotes}
\end{threeparttable}
\end{table}

\subsection{\label{sec:32}Simulation setup}

\begin{figure}[h]
	\centering\includegraphics[width=\columnwidth]{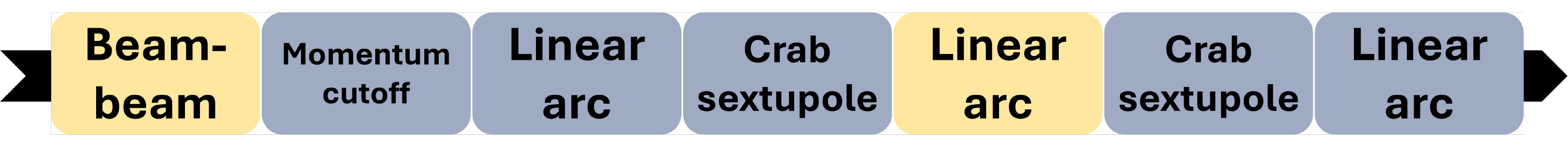}
    \caption{Sequence of elements in the \texttt{Xsuite} tracking setup, starting from the left end, representing one superperiod of the ring. Radiation is modeled in the elements represented with a yellow block.}
    \label{fig:tracking_setup}
\end{figure}

Our tracking model is sketched in Fig.~\ref{fig:tracking_setup}.

It consists of a linear transfer map, which we will refer to as the arc element, representing one superperiod of the FCC-ee ring featuring 4 IPs, with the corresponding betatron and synchrotron tune fractions. We split this element into 3 parts and insert two sextupole elements in between them to implement the crab-waist scheme at the IP. In the middle arc element we update the particle trajectories with an exponential damping and Gaussian noise excitation as an effective implementation of synchrotron radiation. In the beam-beam element we simulate the emission of beamstrahlung photons~\cite{Kicsiny:2887806} and after this element we insert a momentum collimator which cuts off particles which fall out of the momentum acceptance limit. We start each iteration in front of the IP where we record the dynamical variables of the particles at each loop. With this we can obtain the evolution of the r.m.s. bunch sizes over many turns and study the equilibrium values. 

In \texttt{Xsuite} the beamstrahlung is modeled by an algorithm which was originally developed in~\cite{Yokoya:1985ab}, and which is also used in the code \texttt{GUINEA-PIG}~\cite{guineac}, a particle in cell solver used for simulating single beam-beam collisions with background generation. The algorithm in \texttt{Xsuite} was adapted from \texttt{GUINEA-PIG}. In \texttt{Xsuite} the inverse bending radius $1/\rho$ is computed as the ratio of the transverse soft-Gaussian force divided by the longitudinal distance between two consecutive longitudinal slices. The radiation integrals $\mathcal{I}_{n}$ from Eq.~\eqref{eq:2} are not used in the numerical simulation.

In this simulation setup, we track a single pair of colliding bunches (1 bunch per beam) corresponding to a 2 IP machine. In a 4 IP machine, independent sets of 4 bunches (2 bunches per beams) would be colliding with each other. Our setup remains representative of a rather pessimistic yet realistic situation where the two bunches in a given beam feature comparable properties. In case the two bunches in a given beam have different properties, for example due to fluctuations in the injected intensities, the treatment should be revised.

\begin{table}[h!]
\caption{\label{tab:nturns}Number of turns used in our tracking simulations. Each turn consists of 4 iterations over the lattice superperiod, shown in Fig.~\ref{fig:tracking_setup}.}
    \begin{ruledtabular}
    \begin{tabular}{lcccc}
  & \textbf{Z} & \textbf{W$^{\pm}$}& \textbf{ZH} & \textbf{t\=t}\\
    \hline
$N_t$ [1] & 20000 & 10000 & 5000 & 5000 \\

\end{tabular}
\end{ruledtabular}
\end{table}

We performed our studies with 4 FCC-ee parameter sets, shown in Tab.~\ref{tab:FCC}, which are based on recent progress in the machine optics design~\cite{oide_fccweek}. For all simulations we used $N_m=10^6$ macroparticles for the Z mode and $N_m=10^5$ for the other modes, since we observed these numbers to yield statistically converged dynamics. For lower energies more turns are required to reach equilibrium due to the slower synchrotron radiation damping time, which dominates transient dynamics. The number of turns are summarized in Tab.~\ref{tab:nturns}. We count each turn by iterating 4 times through the lattice superperiod, described in the previous section. 

We estimated the optimal number of longitudinal slices in the beam-beam model by~\cite{Shatilov:2816655}
\begin{equation}
\label{eq:nslices}
    N_s=10\cdot\frac{\sigma_{z,\text{BS}}}{\mathrm{min}(L_i,\beta_y^*)},
\end{equation}
where $L_i$ is the interaction length of a collision given by Eq.~\eqref{eq:li}. Both $L_i$ and $\beta_y^*$ are in the order of millimeters. In Eq.~\eqref{eq:nslices} $L_i$ is smaller for Z and ZH, while $\beta_y^*$ is smaller for W$^{\pm}$ and t\=t. The ratio of the bunch length to the waist or interaction length is a measure for the variation of the bunch cross section during collision, with a high ratio indicating that more slices are required for accurate simulation. We used the value of $N_s$ rounded up to the nearest hundred, which resulted in 200 slices for the Z energy and 100 for all other resonances. The longitudinal slicing covers the bunch in the $\pm5\sigma_{z,\text{BS}}$ range. We used the quasi-strong-strong model with an update frequency of the statistical moments in the beam-beam element after every 100 collisions. The equilibrium bunch lengths for all FCC-ee configurations in the symmetric case, simulated this way, agree well with the values reported in Tab.~\ref{tab:FCC}. 

\section{\label{sec:4}Results}

In our first study we scanned the bunch intensity asymmetry by gradually increasing the initial bunch population of the high intensity bunch by $\Delta N$ and decreasing it for the low intensity bunch by the same amount, according to the formula $\allowdisplaybreaks N_{l,h}=N_0(1 \pm \Delta N)$, where $N_0$ represents the nominal, or an "average" bunch intensity from Tab.~\ref{tab:FCC}, and the + (-) sign is to be applied for the high (low) intensity bunch. We limited our scan to the range $\Delta N \in[0-0.2]$ and we omitted the study of larger asymmetries as we are interested in realistic scenarios which can occur during the top-up injection.

\subsection{\label{sec:41}Equilibrium bunch length}

\begin{figure}[h]
    \centering\includegraphics[width=.9\columnwidth]{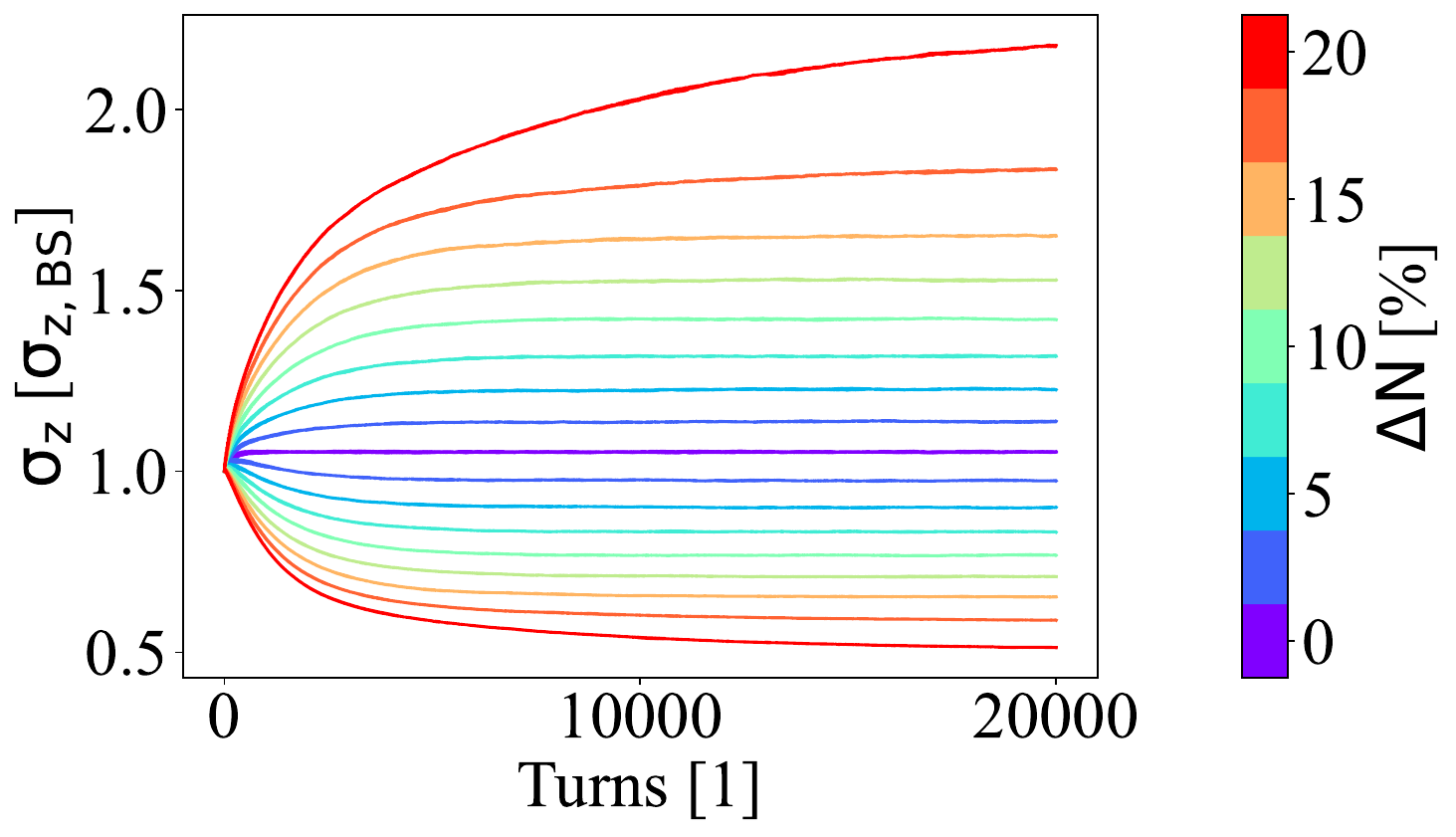}
    \caption{Simulated bunch length for both beams in the FCC-ee Z setup. The colorbar denotes the initial bunch intensity asymmetry. The values are normalized to the nominal equilibrium bunch length from Tab~\ref{tab:FCC}. The low intensity bunch blows up while the high intensity bunch shrinks.}
    \label{fig:sz_evolution}
\end{figure}

Figure~\ref{fig:sz_evolution} shows the turn-by-turn evolution of the r.m.s. bunch length for both beams in the Z operation mode, as a function of the initial bunch intensity asymmetry. The initially low intensity bunch is always the one which blows up, while the initially high intensity bunch shrinks.

\begin{figure}[h]
    \centering\includegraphics[width=.9\columnwidth]{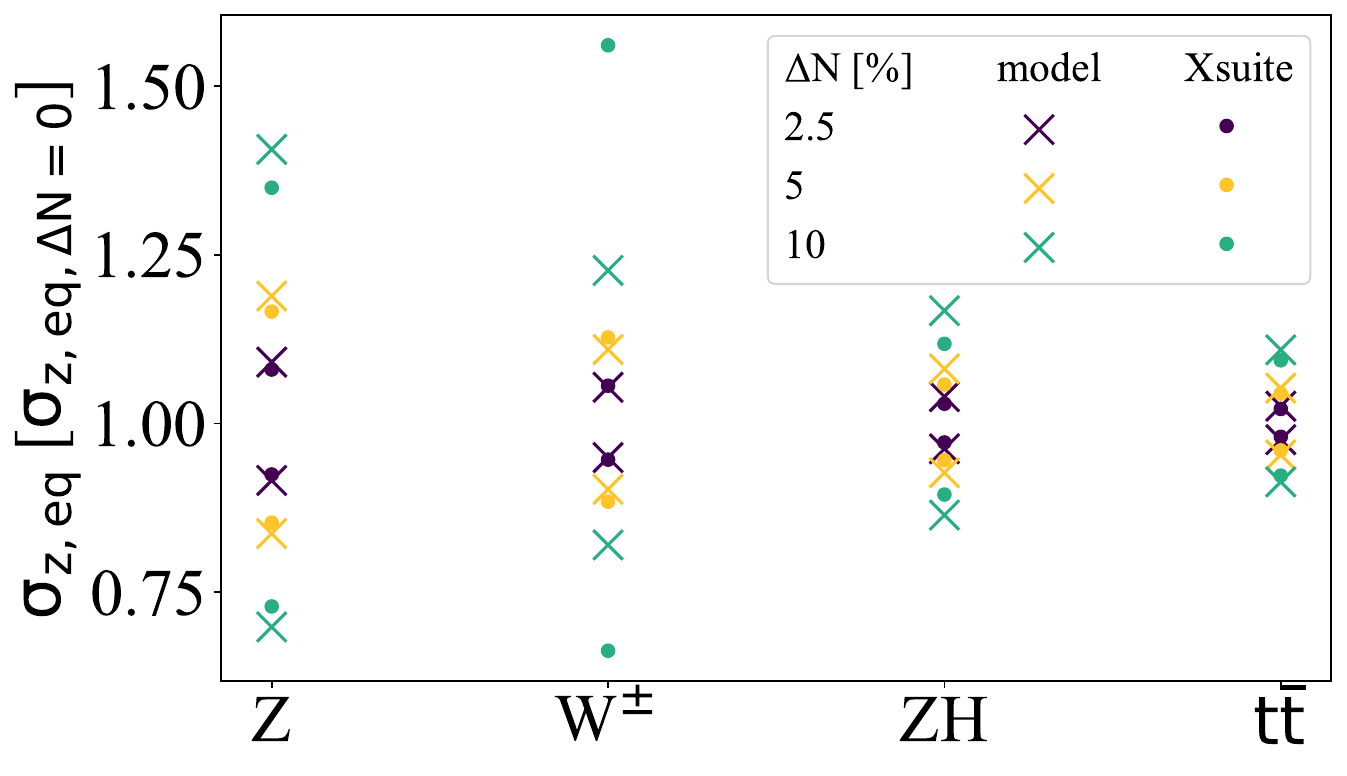}
    \caption{Simulated and predicted equilibrium bunch lengths for the nominal FCC-ee configurations, at selected initial bunch intensity asymmetries. The values are always normalized to the simulated or predicted equilibrium bunch length in the symmetric case, with $\Delta N=0$. The data are calculated from the last 2500 turns, with negligible statistical uncertainties.}
    \label{fig:asy}
\end{figure}

Figure~\ref{fig:asy} shows the equilibrium bunch lengths for all FCC-ee configurations, with different values of the initial bunch intensity asymmetry. The dots show our simulation results with \texttt{Xsuite}, and the crosses indicate the analytical predictions by solving Eq.~\eqref{eq:6}.

From Fig.~\ref{fig:asy} it can be seen that the model predictions and the simulation results are in good agreement. The model becomes inaccurate in the W$^{\pm}$ configuration featuring 10~\% asymmetry. Here the bunch length blows up by more than 50~\%. This is caused by nonlinear diffusion, which at the same time triggers a vertical blowup by an order of magnitude and a horizontal blowup comparable to that in the longitudinal dimension. This 3D flip-flop effect suggests that for certain parameter regimes, especially with high asymmetries and sensitive operation modes, such as the W$^{\pm}$, the 1D model introduced earlier is not accurate, and the behavior cannot easily be captured in a phenomenological model.

\begin{figure}[h]
    \centering\includegraphics[width=.9\columnwidth, scale=0.6]{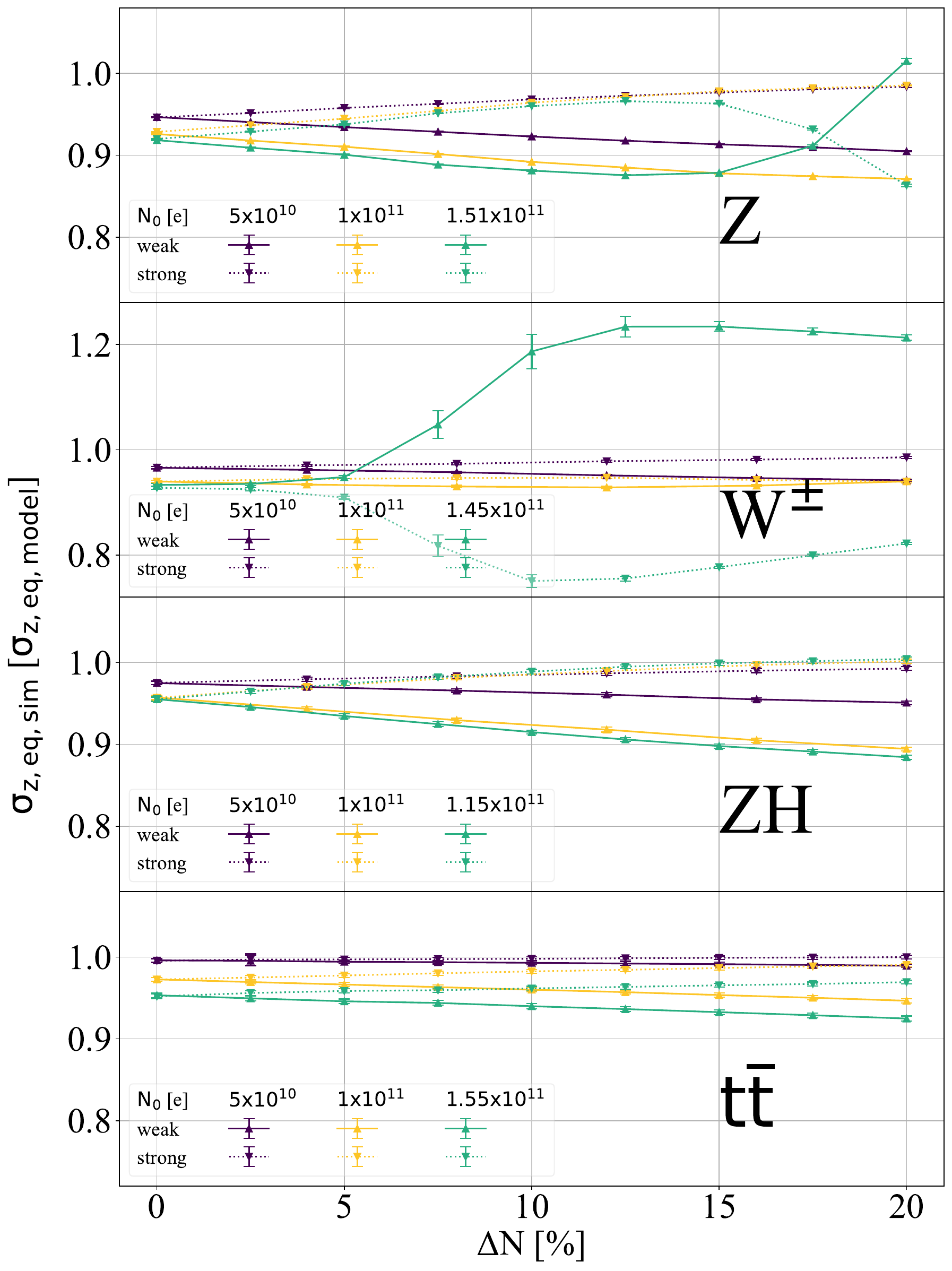}
    \caption{Ratio of simulated to analytically predicted equilibrium bunch lengths for selected average bunch intensities at the FCC-ee. Error bars come from simulation and denote the statistical uncertainty computed from the last 2500 turns.}
    \label{fig:sz}
\end{figure}

The agreement between the simulated and the analytically predicted equilibrium bunch length is further detailed in Fig.~\ref{fig:sz}, showing their ratio for all FCC-ee configurations, with a set of selected average bunch intensities, up to the nominal intensity. Overall, the agreement between the simulation and the model up to a bunch intensity asymmetry of 20~\% is rather good, i.e. within 10~\%. The simulated equilibrium bunch lengths are systematically lower than the model predictions, which could be caused by the approximations done in the analytical model, which, as detailed earlier in Sec.~\ref{sec:2}, assumes small transverse amplitudes, no hourglass and no crab-waist. The main differences arise in configurations much above the onset of strong nonlinear diffusion mechanism, which is the most dominant for the W$^{\pm}$ mode above 5~\% asymmetry at the nominal bunch intensity. We note that in this regime, the particles lost from the perturbed beam due to the stronger beam-beam force and the resulting reduction of the dynamic aperture are neglected in both the analytical and numerical models. In reality, such losses will degrade further the beam quality once the threshold is reached.

\subsection{\label{sec:43}Luminosity}

\begin{figure}[h]
    \centering\includegraphics[width=.9\columnwidth, scale=0.6]{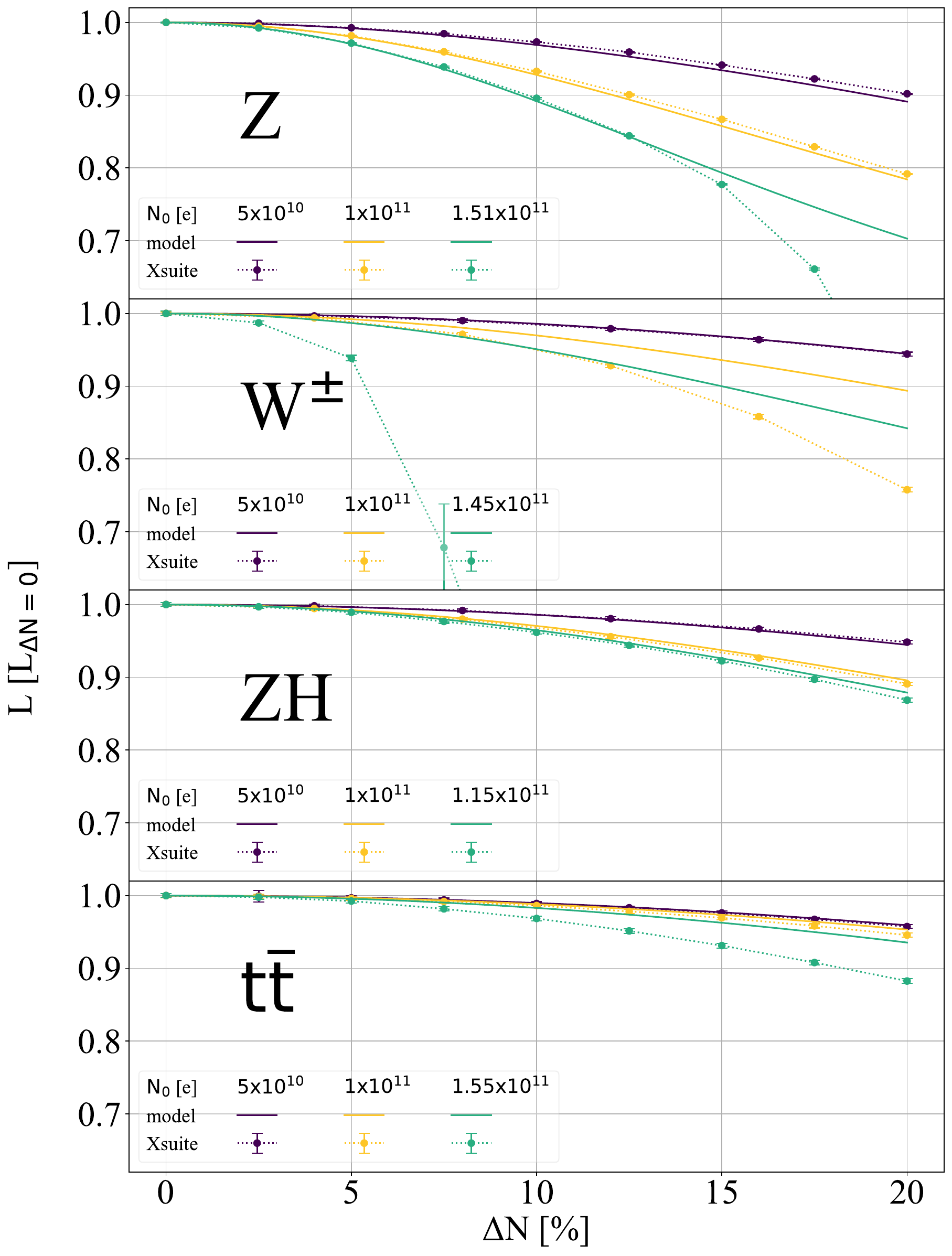}
    \caption{Integrated luminosity of a single collision, averaged from the values computed from the point of view of both beams, at the various FCC-ee resonances, as a function of the initial asymmetry in the bunch population. The values are normalized to the luminosity obtained in the symmetric setup ($\Delta N=0$). Data points indicate the simulation results, calculated from the last 2500 turns. The solid lines show the analytical estimates.}
    \label{fig:lumi}
\end{figure}

The luminosity is expected to decrease in the presence of an intensity asymmetry even without transverse blowup, as the bunch length of the perturbed bunch increases and thus increases the Piwinski angle, as well as because $N_0(1+\Delta N)N_0(1-\Delta N)=N_0^2 (1-\Delta N^2)<N_0^2$. Figure~\ref{fig:lumi} shows the luminosity normalized to the value obtained in the symmetric configuration, as a function of the bunch intensity asymmetry, for all FCC-ee energies and for different average bunch intensities. 

The analytical prediction is made using~\cite{Damerau:905072}:	
\begin{align}
	\label{eqn:lumi}
		\mathcal{L}&=\dfrac{N_lN_h}{T_\text{rev}}\dfrac{\cos(\theta_c/2)}{2\pi}\dfrac{1}{\sqrt{\sigma_{y,l}^{*2}+\sigma_{y,h}^{*2}}}\nonumber\\
		&\times\dfrac{1}{\sqrt{(\sigma_{x,l}^{*2}+\sigma_{x,h}^{*2})\cos^2(\theta_c/2)+(\sigma_{z,l}^2+\sigma_{z,h}^2)\sin^2(\theta_c/2)}},
\end{align}

where we use the nominal transverse bunch sizes from Tab.~\ref{tab:FCC} and the equilibrium bunch lengths by solving Eq.~\ref{eq:6}. The overall agreement is again good. In the Z and W$^{\pm}$ configurations we observe increasing discrepancies for configurations featuring a high average intensity and high asymmetry due to the fact that the 1D model is not accurate enough much above the onset of the 3D flip-flop. 

By using the analytical model we have validated our numerical simulation for the physics of the flip-flop mechanism in the longitudinal plane. For the transverse planes, a similar analytical model is hard to develop due to the highly nonlinear dynamics, therefore this can only be studied reliably with numerical simulations. In the next section, we use the same simulation setup to make predictions on the equilibrium dynamics in the top-up injection under different asymmetric starting conditions.

\section{\label{sec:5}Top-up injection}

The FCC-ee is planned to operate in the so called top-up injection scheme~\cite{Aiba:2815853}. There are two main variants of this technique, namely on-momentum off-axis (transverse) and off-momentum on-axis (longitudinal) top-up injection~\cite{Ramjiawan:2022fco}. In case of off-axis injection, the injected beam is offset from the stored beam typically at some distance in the horizontal direction. This option is currently less favored due to the difficulties to find a suitable collimator setup, which is able to absorb synchrotron radiation emitted from the injected bunch and the stored bunch at the same time~\cite{Hofer:2845885}. We therefore focus on the on-axis injection scheme for which the low intensity bunches are injected with a momentum offset.

In the following parts, we first present our study of configurations with a fixed asymmetry in the initial bunch population and vary the injection offset. Next, we investigate the impact of the bunch intensity asymmetry and finally the asymmetry in the beam size.

\subsection{Simulation of longitudinal top-up injection}

\begin{figure}[h]
	\centering\includegraphics[width=\columnwidth]{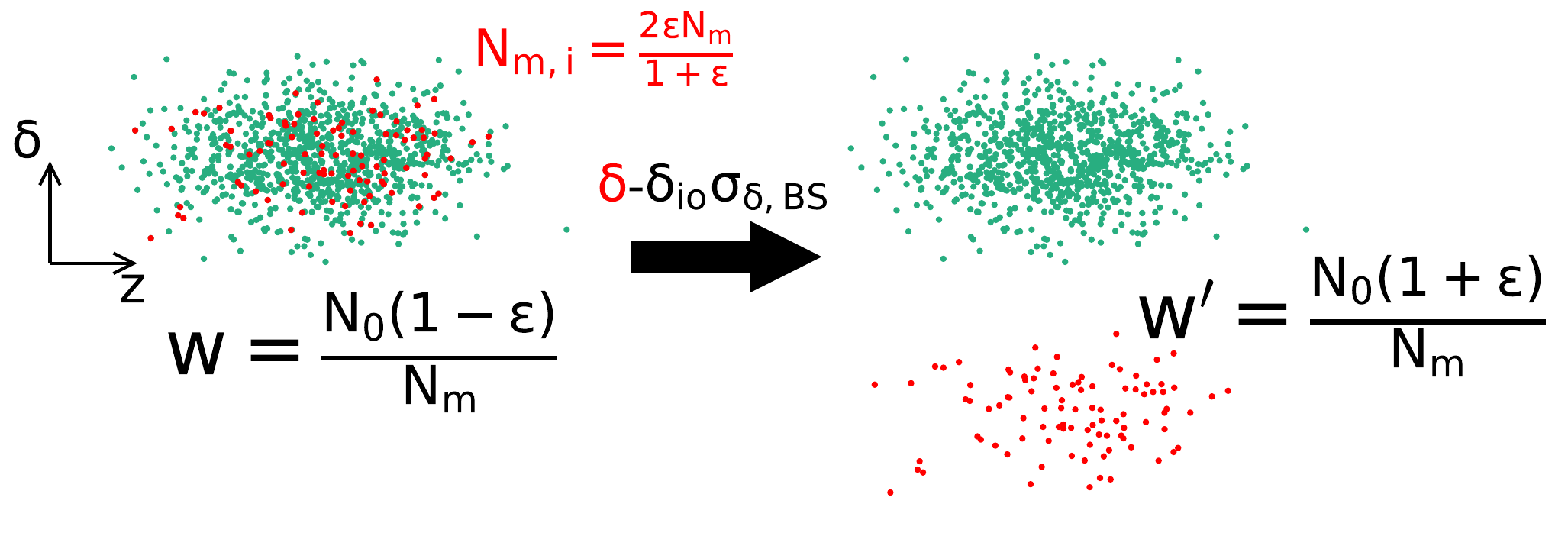}
    \caption{Longitudinal phase space of the perturbed bunch just before (left) and after (right) longitudinal top-up injection as modeled in \texttt{Xsuite}. Injection is done by offsetting the relative energy of $N_{m,i}$ not yet lost macroparticles by $\delta_{\text{io}}\sigma_{\delta, \text{BS}}$ and updating all weights from $w$ to $w'$.}
    \label{fig:topup_sim}
\end{figure}

The dynamic aperture can affect top-up injection efficiency, i.e. the number of survived particles. We investigated this tolerance in the presence of asymmetric beam-beam interactions by using a linear lattice model with the beam-beam element, including beamstrahlung and a "hard-edge" dynamic aperture limit that implements a cutoff above and below a certain threshold, corresponding to the momentum acceptance. From previous beam-beam simulations, the tolerance for the bunch intensity difference at top-up injection was set to $\pm$5~\%~\cite{Benedikt:2651299}. We simulate a scenario in which one of the beams (the $e^+$) has lost 5~\% of the nominal bunch intensity (perturbed bunch), while the second beam (the $e^-$) is kept at its nominal intensity (fixed bunch). That is, we initialize the fixed bunch with $N_0$ and the perturbed bunch with $N_0(1 - \varepsilon)$, where $\varepsilon=0.05$. With this setup we tracked the two beams to let them converge to an equilibrium. In this first stage we used the quasi-strong-strong model with an update frequency of the statistical moments in the beam-beam element after every 100 collisions. Due to the flip-flop effect, the equilibrium bunch sizes are expected to be asymmetric, but finite. When the equilibrium is reached, we top-up the perturbed bunch to an intensity of $N_0(1 + \varepsilon)$ which is representative to a real life scenario. In a real machine the injection will always overshoot the target intensity by a small margin to allow for the intensity to decay until the next injection. In this second stage, following the intensity change of the perturbed bunch, the beams are tracked again to observe the converged beam profile after injection. Here we used the full strong-strong model as the merging of the injected bunch into the stored bunch leads to a turn-by-turn change in the beam-beam force. The procedure of topping up the perturbed bunch is sketched in Fig.~\ref{fig:topup_sim}. 

Both beams are initialized with $N_m$ macroparticles. The initial weight of the perturbed bunch particles is set to $w=N_0(1 - \varepsilon)/N_m$. At the top-up, the weights of all surviving particles in the perturbed bunch are simply updated to $w'=N_0(1 + \varepsilon)/N_m$. Then a random subset of the (not yet lost) perturbed bunch macroparticles is selected, corresponding to a bunch intensity of $2\varepsilon N_0$. This corresponds to a fraction $2\varepsilon / (1 + \varepsilon)$ of the initial number of macroparticles, which are then offset by a relative energy of $-\delta_{\text{io}}$, called the injection offset. The two beams are then tracked until the equilibrium is reached. The number of tracking turns in both stages is equal to the value shown in Tab.~\ref{tab:nturns}.

\begin{figure}[h]
	\centering\includegraphics[width=.9\columnwidth]{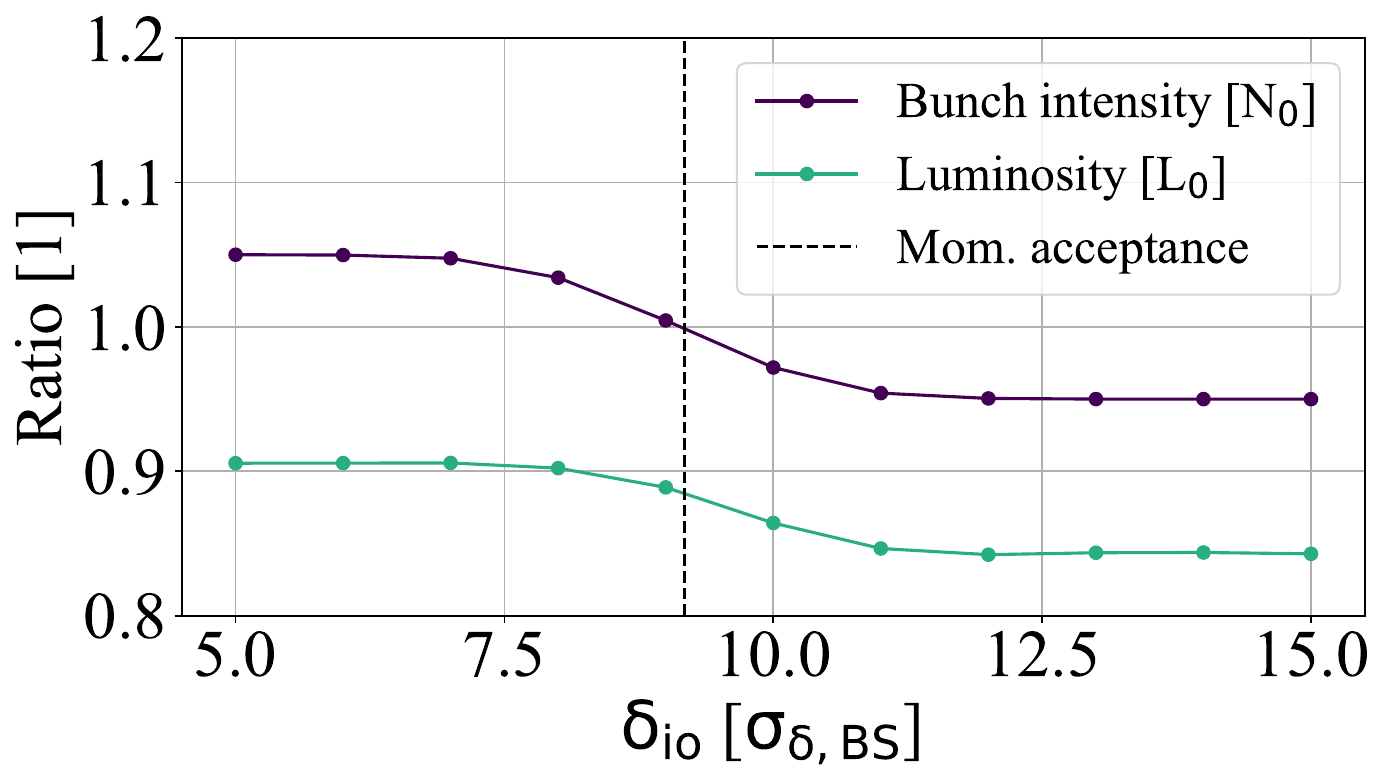}
    \caption{Bunch intensity of the perturbed bunch, normalized to the nominal value from Tab.~\ref{tab:FCC}, and luminosity, averaged from the values computed from the point of view of both beams, normalized to the simulated luminosity $L_0$ in the symmetric setup from a single collision in the strong-strong model. All data are calculated from the last 2500 turns, after top-up injection, as a function of the injection offset $\delta_{\text{io}}$, for the W$^{\pm}$ operation mode, with a momentum acceptance of 1~\%.}
    \label{fig:topup_w}
\end{figure}

The typical fraction of bunch intensity and luminosity which remains after the top-up injection as a function of the injection offset $\delta_{\text{io}}$ is shown in Fig.~\ref{fig:topup_w} for the W$^{\pm}$ operation mode, which is the most sensitive to the flip-flop effect, therefore it provides a baseline for discussion. 

\begin{figure}[h]
	\centering\includegraphics[width=\columnwidth]{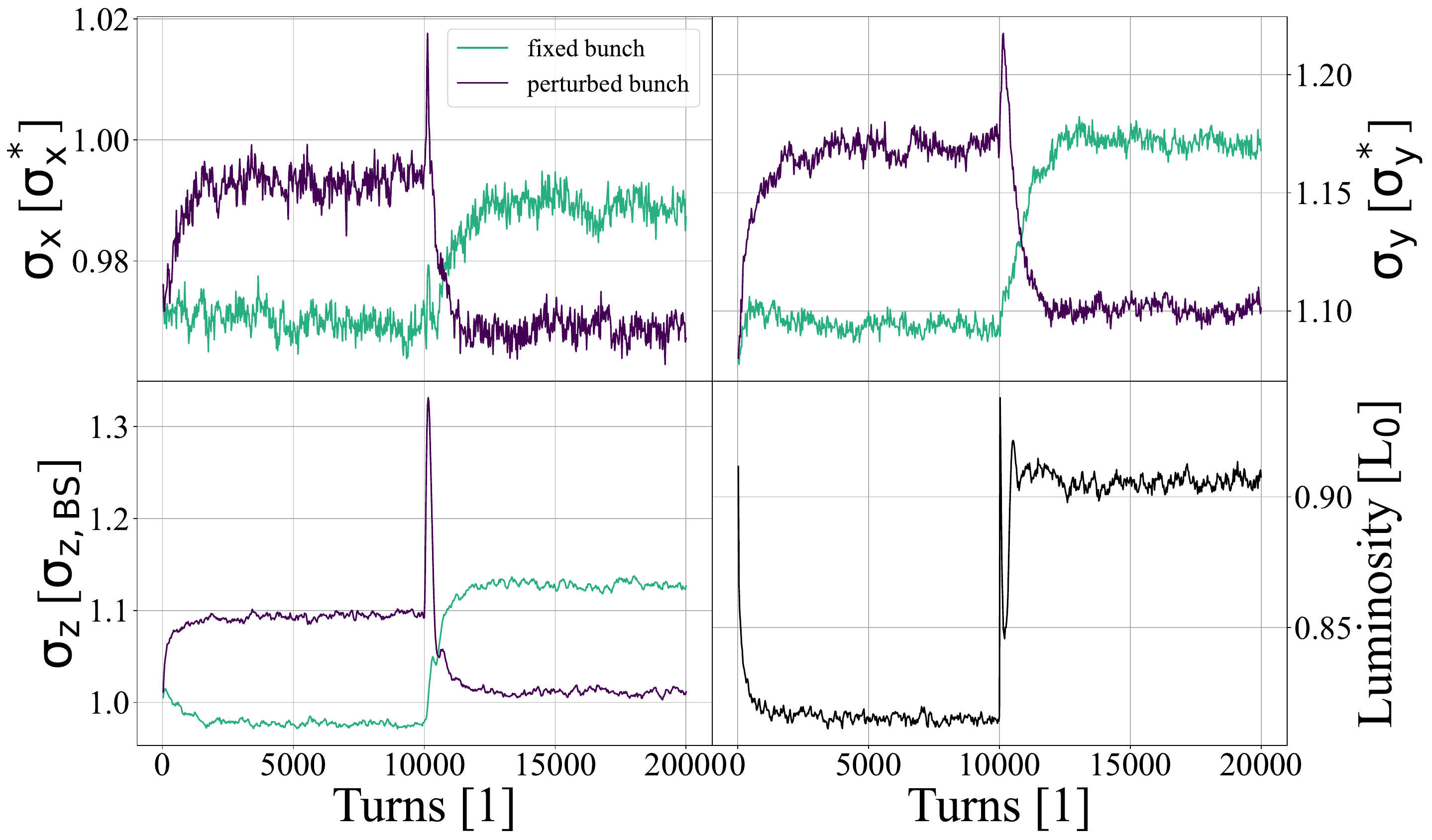}
    \caption{Evolution of r.m.s. bunch sizes, normalized to their nominal values from Tab.~\ref{tab:FCC}, and luminosity, averaged from the values computed from the point of view of both beams, normalized to the simulated luminosity $L_0$ in the symmetric setup from a single collision in the strong-strong model. All data are for for the W$^{\pm}$ mode in a simulation of the longitudinal top-up injection, using $\delta_{\text{io}}=7$, with a momentum acceptance of 1~\%.}
    \label{fig:topup_w_ev}
\end{figure}

We observed that the replenished bunch loses a fraction of the injected intensity at the moment of the injection, depending on $\delta_{\text{io}}$. We note that a trivial limit to $\delta_{\text{io}}$ is the momentum acceptance, which is highlighted in Fig.~\ref{fig:topup_w}. At this nominal momentum acceptance limit, approximately half the injected intensity is lost. The full intensity is successfully retained with a momentum acceptance less than $7\sigma_{\delta,\text{BS}}$. The injection is followed by a transient phase of merging with the stored beam, visible in Fig.~\ref{fig:topup_w_ev} as a peak most dominant in the vertical and longitudinal direction. The nontrivial question was whether this transient behavior could lead to a permanent degradation of the luminosity by triggering a transverse-longitudinal flip-flop mechanism.

After the transient, we found that the luminosity stabilizes to a higher equilibrium value, as the average bunch intensity is higher than before injection. Yet we note that the overall luminosity is lower than its design value due to the asymmetry in intensity and bunch length which remains all along the process. In a sense, the perturbed and fixed bunches have exchanged their role following the injection, but the asymmetry remains. While setting the injection offset to be larger than the momentum acceptance results in a sudden drop in the luminosity, there are no unexpected transients leading to additional losses for injections with momentum offsets above the trivial threshold. Based on our results, the transient behavior due to longitudinal top-up injection, i.e. the merging of the injected bunch into the stored bunch, does not trigger the flip-flop mechanism to an extent larger than predicted by a steady-state simulation, i.e. one without the injection process where the bunches are initialized with a given bunch intensity asymmetry. Moreover, the equilibrium bunch length and vertical beam size are in each case within $\pm$20~\% compared to the nominal equilibrium value.

\subsection{Dependence on the bunch intensity asymmetry}

\begin{figure}[h]
    \centering\includegraphics[width=.9\columnwidth]{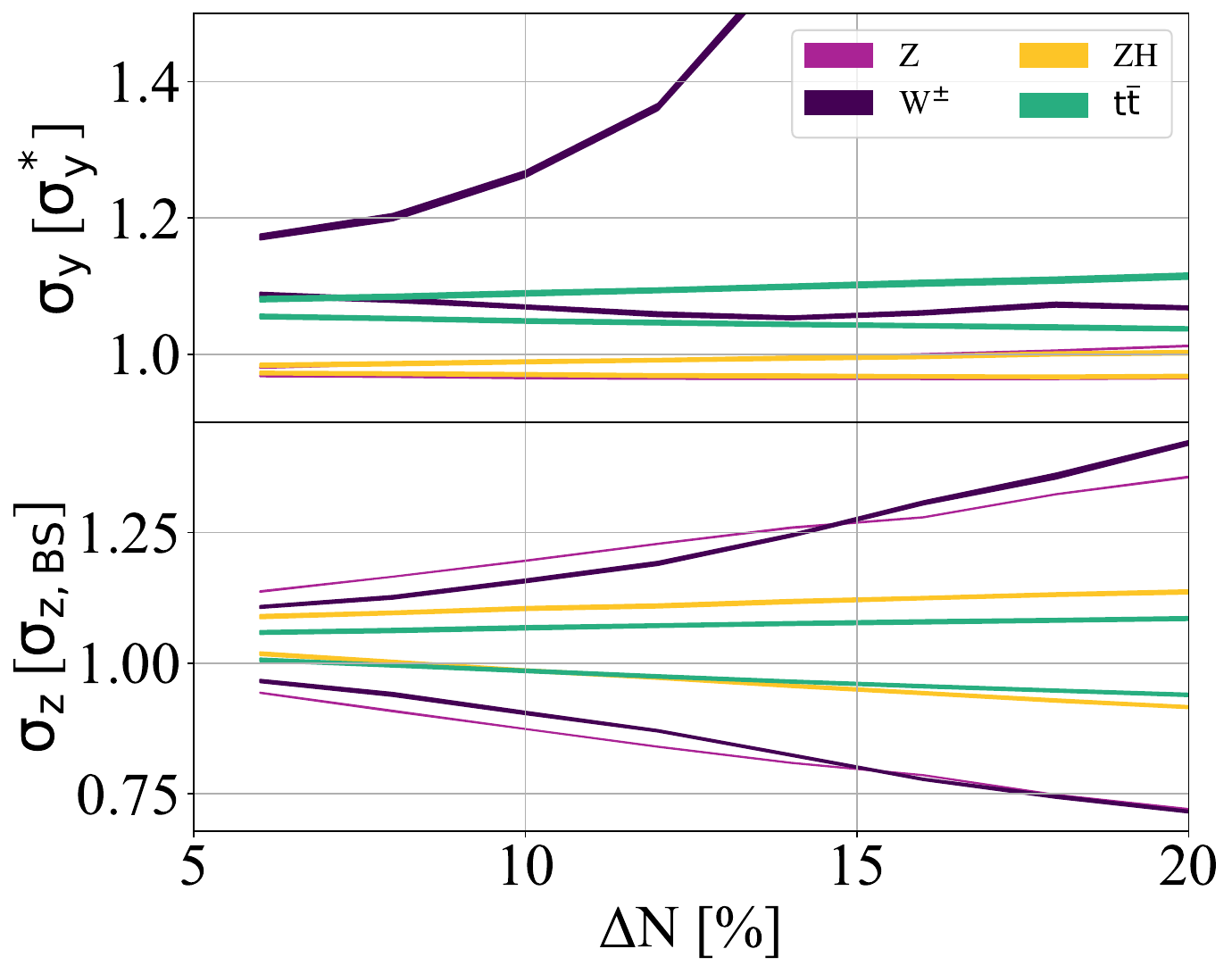}
    \caption{Equilibrium vertical r.m.s. bunch size (top) and bunch length (bottom) before injection, calculated from the last 2500 turns, as a function of the initial bunch intensity asymmetry for all FCC-ee operation modes, simulated with their respective nominal momentum acceptance.}
    \label{fig:topup2}
\end{figure}

In the previous study it was assumed that the beam is topped up when its bunch intensity drops by 5~\%. This top-up threshold asymmetry is however subject to optimization in a real machine, therefore it is important to study the equilibrium dynamics in cases with potentially higher values. In a subsequent study we performed a scan of the initial bunch intensity asymmetry, by increasing it gradually from $\Delta N = 6~\%$ up to 20~\%. In this setup, the perturbed bunch intensity is given by $\allowdisplaybreaks N_{w}=N_0(1- \Delta N)$, while the fixed bunch intensity is kept at $N_0$. We used the nominal momentum acceptance for each operation mode, shown in Table~\ref{tab:FCC}. Figure~\ref{fig:topup2} shows the equilibrium vertical bunch size and bunch length values, recorded after reaching equilibrium, before injection, i.e. after tracking for a number of turns shown in Tab.~\ref{tab:nturns}. 

We have observed that for the lower energies (Z and W$^{\pm}$) the equilibrium r.m.s. bunch length in particular is more sensitive to the initial asymmetry, suggesting an inverse relation with the beam energy. Regarding the vertical r.m.s., we observed the onset of the transverse-longitudinal flip-flop already around 10~\%. 

\begin{figure}[h]
    \centering\includegraphics[width=.9\columnwidth]{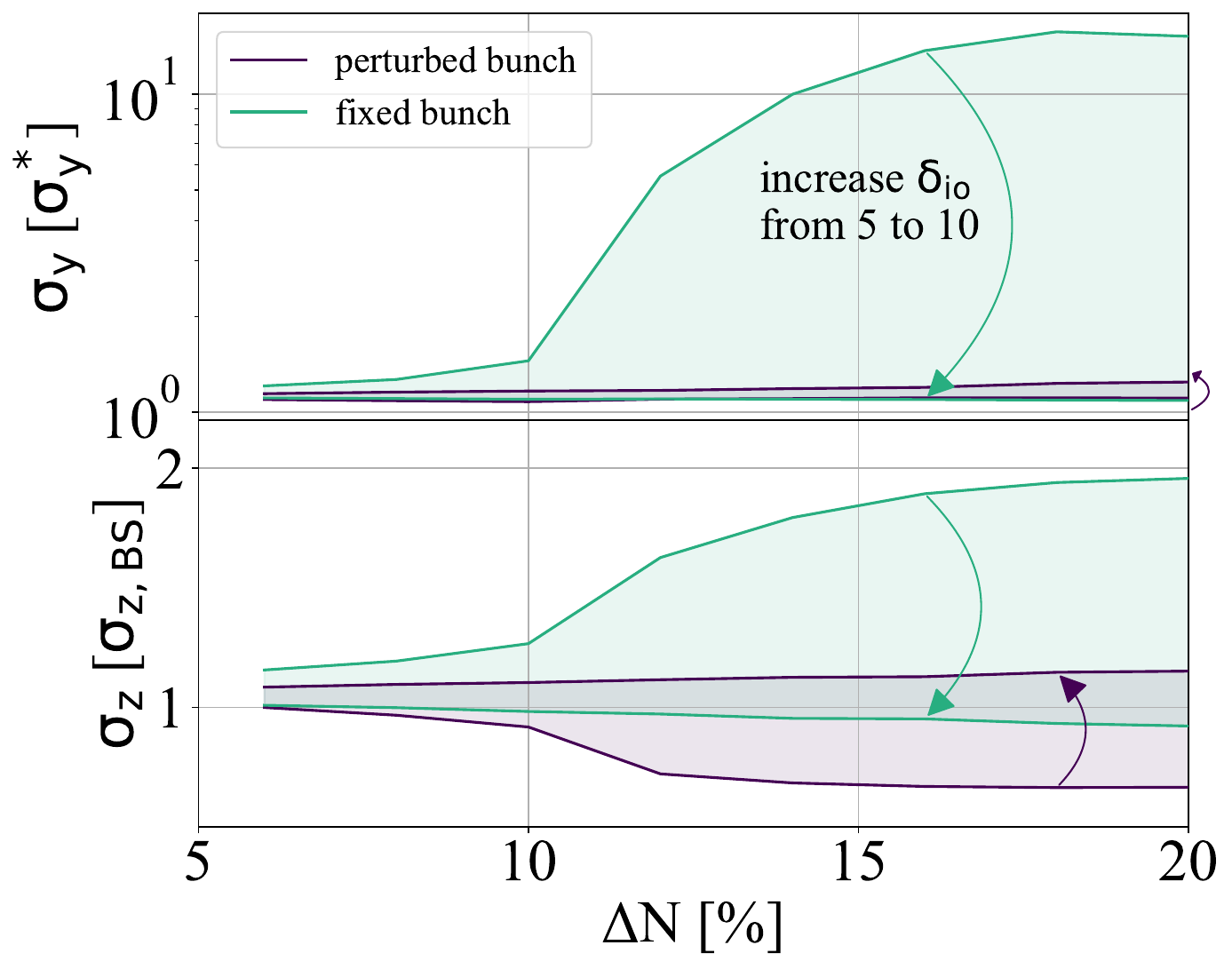}
    \caption{Equilibrium vertical r.m.s. bunch size (top) and bunch length (bottom) after injection, calculated from the last 2500 turns, as a function of the initial bunch intensity asymmetry for the FCC-ee W$^{\pm}$ mode, simulated with its nominal momentum acceptance. The arrows show the change in the equilibrium when increasing $\delta_{\text{io}}$ from 5 to 10 in units of $\sigma_{\delta,\text{BS}}$.}
    \label{fig:topup3}
\end{figure}

We have investigated the equilibrium dynamics after performing the top-up injection of the perturbed bunch and by applying two different $\delta_{\text{io}}$ injection offsets (5 and 10 $\sigma_{\delta,\text{BS}}$ respectively). We show our results for the W$^{\pm}$ resonance in Fig.~\ref{fig:topup3}. We observed that a higher injection offset decreases the asymmetry in the equilibrium beam sizes for Z and W $^{\pm}$. For ZH and t\=t we observed negligible impact on the injection offset. In general, the onset of the vertical flip-flop for W$^{\pm}$ takes place at a smaller bunch intensity asymmetry than prior to the top-up injection. This can be explained by the increased total bunch intensity after the injection, which increases the beam-beam force.

\section{Lattice parameter asymmetry} \label{sec:51}

As discussed earlier, the magnitude of the blowup caused by the flip-flop effect has a dependence on the intensity of beamstrahlung, which is directly related to an asymmetry in the beam-beam force. This asymmetry can result not only from an initial difference in the bunch currents but also from differences in the optical beta functions at the IP or in the equilibrium lattice emittances, which are properties of the lattice. In a real machine, such a situation could occur after optics corrections, where due to the imperfection of this process, some residual imperfections will always remain in the corrected optics. This is particularly important for the vertical emittance, since it is entirely defined by the coupling correction. A slight decrease (increase) in the spot size at the IP results in the increase (decrease) of the density of particles, which results in the same effect as increasing (decreasing) the bunch current but keeping the spot size the same. Therefore it is possible that an initial asymmetry in the beam sizes, resulting from an imperfect matching of the emittances or beta functions in the two rings, can potentially trigger the flip-flop mechanism. In order to avoid such scenarios it is important to give estimates on the tolerances for the error on the optics parameters. 

In the subsequent study we investigated this by introducing an asymmetry in the initial emittances and optical $\beta^*$ functions. We always keep the size of the $e^-$ bunch (called fixed bunch) at its nominal value, and scan the parameters of the $e^+$ bunch (called perturbed bunch), such that in all configurations
\begin{equation}
\begin{split}
     u_+ &= \zeta u_0, \\   
     u_- &= u_0,
\end{split}
\label{eq:uscan}
\end{equation}
where the subscripts $+/-$ stand for the positron and electron beams respectively. Furthermore, we introduced the $\zeta$ unitless scaling factor and $u\in\{\varepsilon, \beta^*\}$. When tuning the emittance, the corresponding lattice equilibrium is also changed. When tuning $\beta^*$, the crab sextupole strength $k_2$ is also adjusted correspondingly. With this setup we performed tracking, with the number of turns indicated in Tab.~\ref{tab:nturns}, to observe the equilibrium bunch profile. We recorded the r.m.s. bunch sizes in front of the beam-beam element in Fig.~\ref{fig:tracking_setup}. Without loss of generality, we present here our results obtained for the W$^{\pm}$ resonance, as this one proved to be the most sensitive to asymmetries based on our previous studies. 

When scanning one parameter, the others are always kept at the nominal value. The fixed bunch is always initialized with the nominal parameters. When we present normalized quantities in the following, the normalization factor of the perturbed bunch changes accordingly with the scanned parameter, since $\sigma^*=\sqrt{\varepsilon\beta^*}$. We present our plots for the range $\zeta\in[0.1-2]$ to explore a wide range of dynamics, however in real life scenarios an imperfection resulting from matching is unlikely to be that large. In general, our results highlight the beam dynamics under the perturbation of the spot size.

\begin{figure}[h]
    \centering\includegraphics[width=.9\columnwidth]{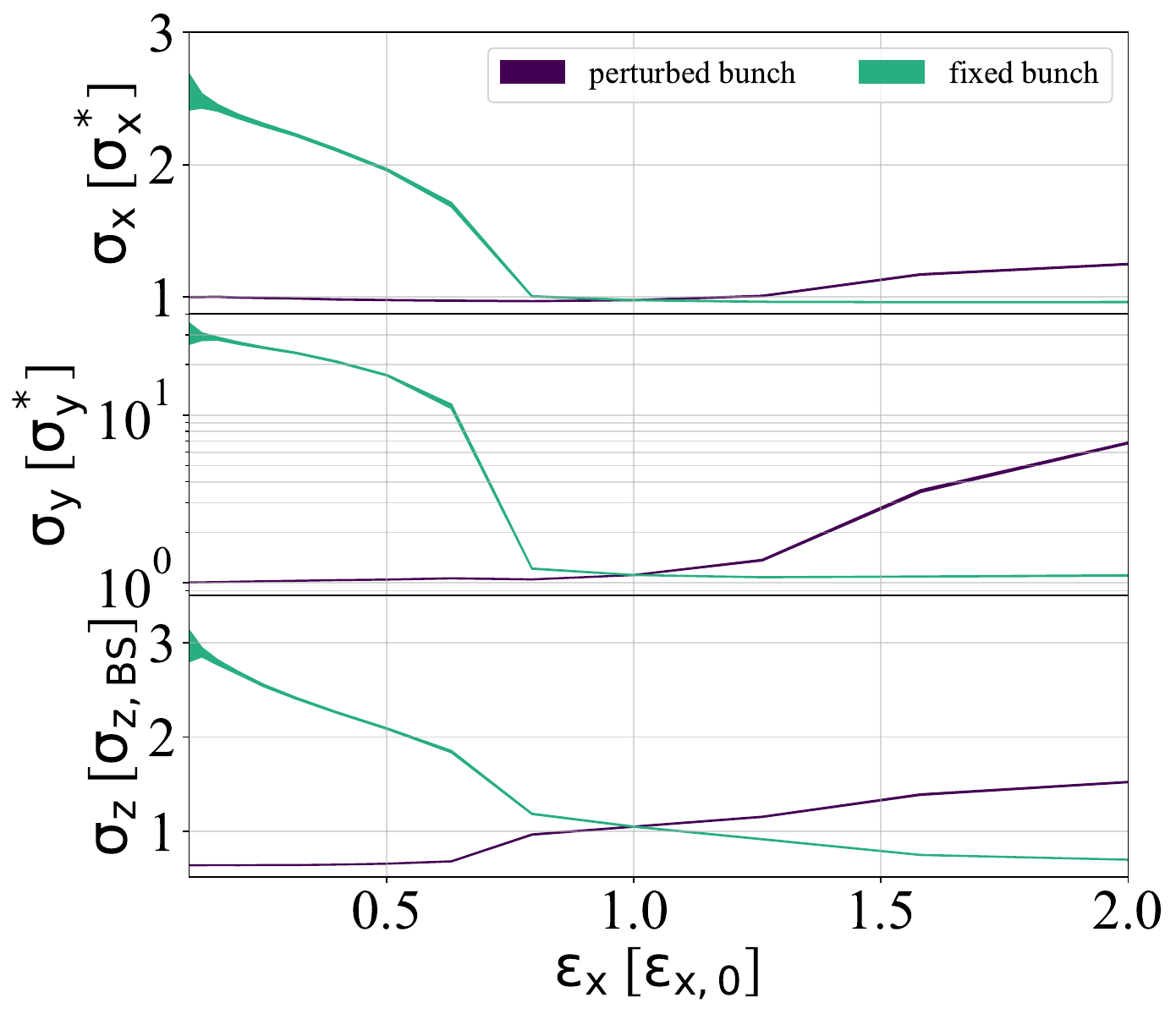}
    \caption{Equilibrium r.m.s. bunch sizes, calculated from the last 2500 turns, as a function of $\varepsilon_{x}$ of the perturbed bunch, in units of their nominal equilibrium value. The fixed bunch is always kept at the nominal parameters, shown in Tab.~\ref{tab:FCC}.}
    \label{fig:beamsize_ex}
\end{figure}

\begin{figure}[h]
    \centering\includegraphics[width=.9\columnwidth]{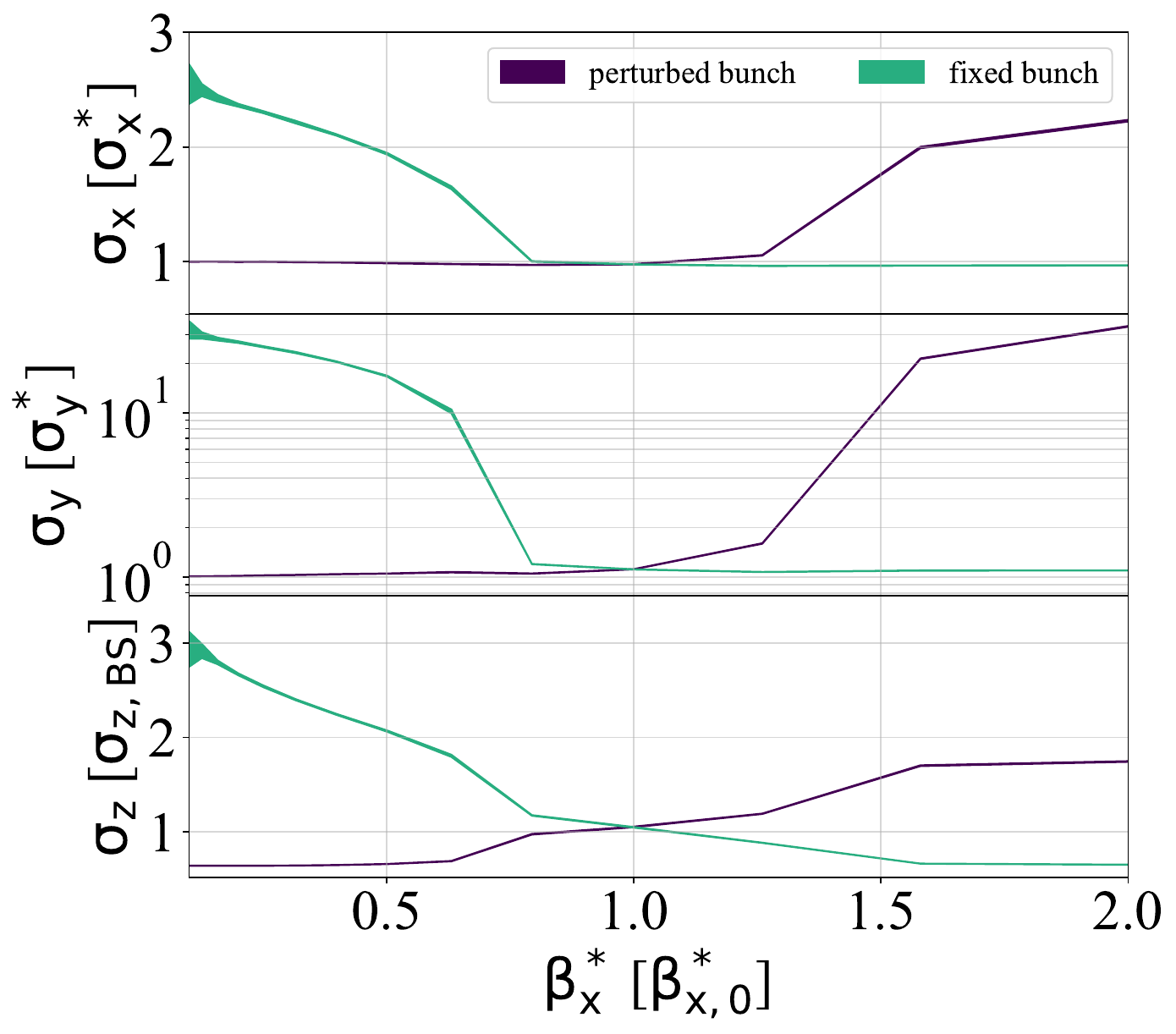}
    \caption{Equilibrium r.m.s. bunch sizes, calculated from the last 2500 turns, as a function of $\beta^*_{x}$ of the perturbed bunch, in units of their nominal equilibrium value. The fixed bunch is always kept at the nominal parameters, shown in Tab.~\ref{tab:FCC}.}
    \label{fig:beamsize_bx}
\end{figure}

Figures~\ref{fig:beamsize_ex} and~\ref{fig:beamsize_bx} show the equilibrium r.m.s. for the two bunches as a function of the horizontal beam parameters $\varepsilon_{x}$ and $\beta_{x}^*$ of the perturbed bunch, normalized to the nominal equilibrium r.m.s. bunch sizes. 

\noindent
The plots are not symmetric with respect to $\zeta=1$ as decreasing beam sizes ($\zeta<1$) tend to increase the strength of the beam-beam force thus making the beams more sensitive to asymmetries. There is no significant difference in the behavior when decreasing $\beta_x^*$ or $\varepsilon_x$, both result in similar equilibrium r.m.s. sizes in all directions. When the beam size of the perturbed bunch increases, its charge density decreases which makes it act with a lower beam-beam force on the opposing bunch, which will therefore undergo less intense beamstrahlung and allows this bunch to converge to a final equilibrium which is closer to the lattice value, without beamstrahlung. This in effect makes the perturbed bunch blow up more. The dynamics is opposite when we decrease the beam size of the perturbed bunch, thereby making it more dense. This will switch the role of the two bunches and the opposing bunch will now experience a stronger beam-beam force. Consequently, it will undergo more beamstrahlung resulting in a blowup, whereas the perturbed bunch will now experience a shrinking. In general, the equilibrium is slightly more sensitive to negative errors in the horizontal parameters.

\begin{figure}[h]
    \centering\includegraphics[width=.9\columnwidth]{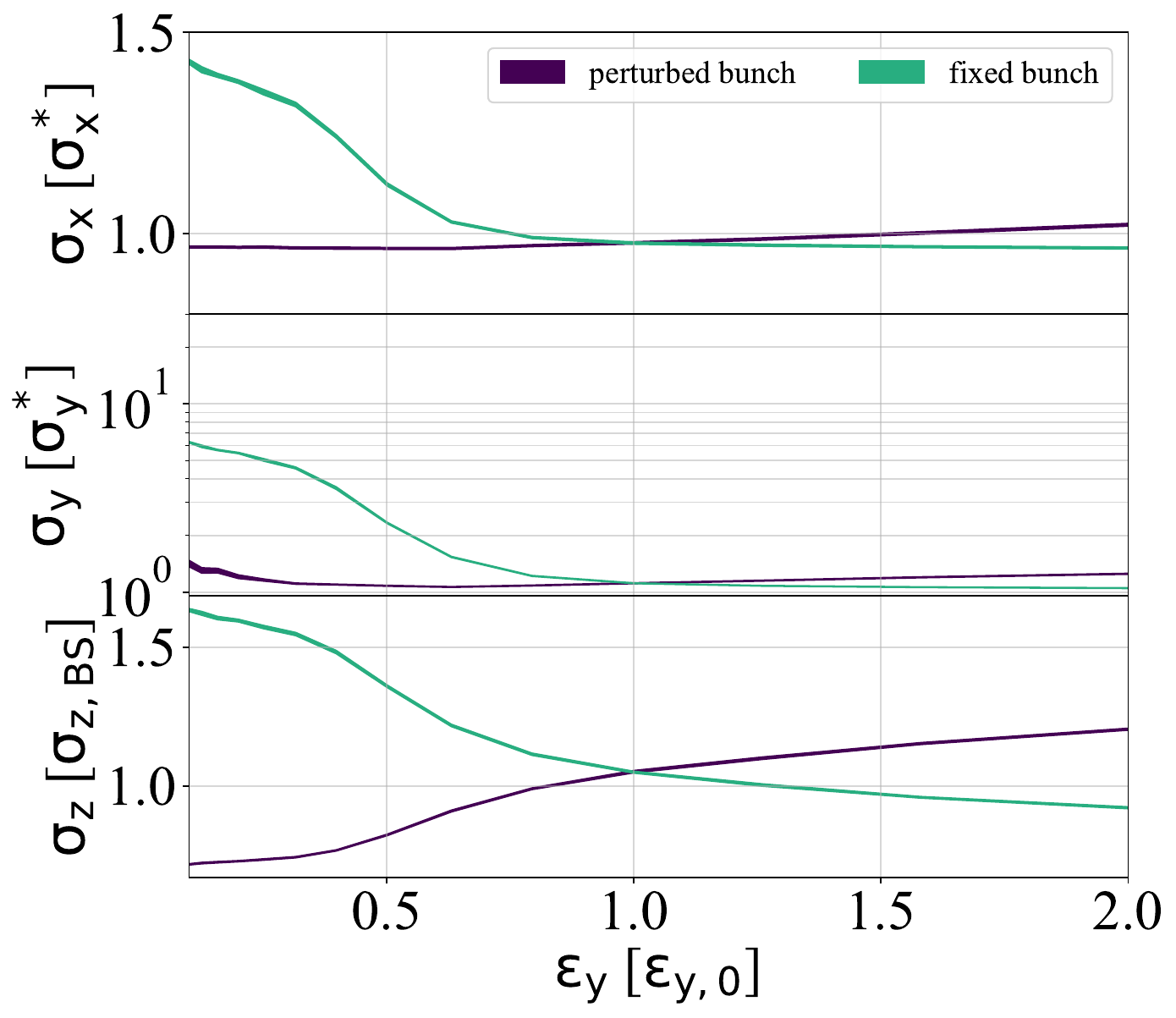}
    \caption{Equilibrium r.m.s. bunch sizes, calculated from the last 2500 turns, as a function of $\varepsilon_{y}$ of the perturbed bunch, in units of their nominal equilibrium value. The fixed bunch is always kept at the nominal parameters, shown in Tab.~\ref{tab:FCC}.}
    \label{fig:beamsize_ey}
\end{figure}

\begin{figure}[h]
    \centering\includegraphics[width=.9\columnwidth]{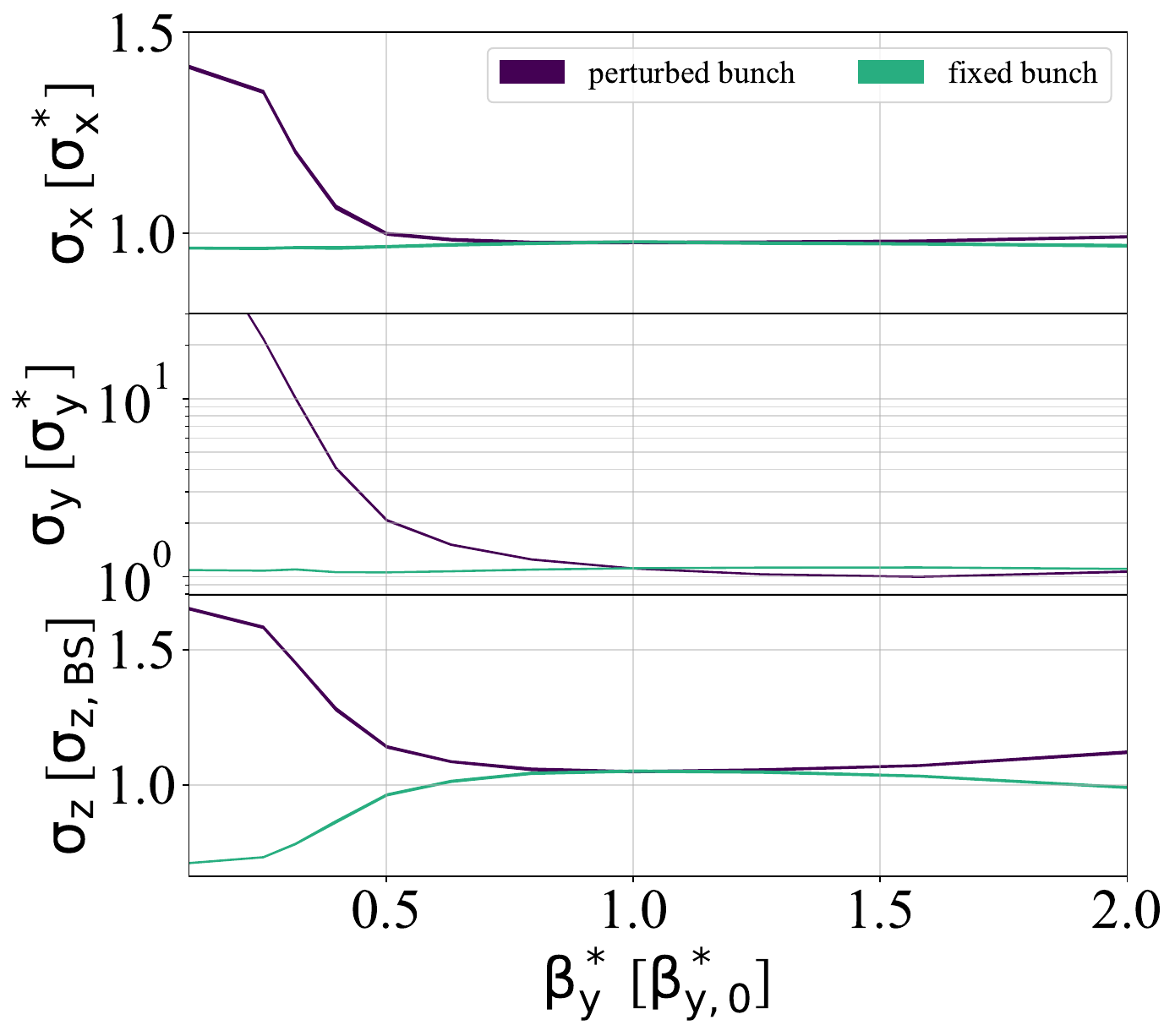}
    \caption{Equilibrium r.m.s. bunch sizes, calculated from the last 2500 turns, as a function of $\beta^*_{y}$ of the perturbed bunch, in units of their nominal equilibrium value. The fixed bunch is always kept at the nominal parameters, shown in Tab.~\ref{tab:FCC}.}
    \label{fig:beamsize_by}
\end{figure}

Figures~\ref{fig:beamsize_ey} and~\ref{fig:beamsize_by} show the equilibrium r.m.s. bunch sizes as a function of the initial vertical beam parameters of the perturbed bunch, in the range of factor [0.1-2]. 

\noindent As for the horizontal plane, the lower beam sizes are more critical than the higher beam sizes, yet the behavior with respect to the emittance and $\beta^*$ differ significantly. Decreasing the vertical emittance of the perturbed bunch results in the expected behavior, by making it more dense and therefore causing the fixed bunch to blow up more. On the contrary, decreasing $\beta_y^*$ causes the perturbed bunch to blow up, albeit becoming more dense. This suggests that with a smaller $\beta^*_y$ the perturbed bunch will undergo more intense beamstrahlung, and the increase in its charge density is compensated by another effect, which in turn reverses the role of the 2 bunches. This additional effect is likely an additional nonlinear diffusion arising from the fact that the crab-waist is sub-optimal with asymmetric optics. Such an effect would deserve detailed studies beyond the scope of this paper.

There is a slight difference in the behavior of the perturbed bunch equilibrium, depending on the tuning of $\beta_x^*$ or $\varepsilon_x$. Increasing $\beta_x^*$, as compared to $\varepsilon_x$, has a stronger effect on the perturbed bunch blowup while it has negligible difference in the equilibrium of the fixed bunch. One might think that this is linked to the change in the crab sextupole strength, defined as:
\begin{equation}
    k_2 = \sqrt{\frac{\beta^*_x}{\beta_{x,s}}}\frac{1}{\theta_c\beta^*_y\beta_{y,s}},
\end{equation}
with the subscript $s$ denoting the optical functions at the location of the crab sextupole. It can be seen that $k_2$ is proportional to $\sqrt{\beta_x^*}$. On one hand, a higher value of $k_2$ means aligning the perturbed bunch waist better against the longitudinal axis of the opposing bunch. On the contrary, if $k_2$ is not adjusted together with $\beta_x^*$, we end up with a less optimal crab-waist which increases transverse blowup. We have repeated the $\beta^*$ scan without changing $k_2$ to see which of these effects dominates the beam dynamics. When increasing $\beta_x^*$, and by keeping $k_2$ unchanged, we found negligible change in $\sigma_{x,\text{eq}}$ and in $\sigma_{z,\text{eq}}$, and 10~\% increase in $\sigma_{y,\text{eq}}$. When scanning $\beta_y^*$, we found up to 50~\% bigger blowup in $\sigma_{x,\text{eq}}$ and in $\sigma_{z,\text{eq}}$, and up to about factor 3 bigger blowup in $\sigma_{y,\text{eq}}$. This can be explained by the suboptimal setting of the crab-waist strength. However, the trend in the blowup as a function of $\zeta$ is still the same as when $k_2$ is adjusted, and different from the one obtained by scanning $\varepsilon_x$. This different trend can better be understood with the horizontal beam-beam parameter $\xi_x$, which is commonly used to characterize the linearized beam-beam force strength experienced by a bunch in this plane. It is a function of the bunch parameters and can be approximated for flat beams ($\sigma_y^*\ll \sigma_{x,\text{eff}}^*$) as:
\begin{align}
    \begin{split}
    \xi_{x,p/f}& \sim \frac{\beta_{x,p/f}^*}{\sigma_{x,\text{eff},f/p}^*(\sigma_{x,\text{eff},f/p}^* + \sigma_{y,f/p}^*)} \approx \dfrac{\beta_{x,p/f}^*}{\sigma_{x,\text{eff},f/p}^{*2}} \\
    & = \frac{\beta_{x,p/f}^*}{\beta_{x,f/p}^*\varepsilon_{x,f/p} + \sigma_z^2\tan^2(\theta_c/2)},\\
    \label{eqn:bbparams}
    \end{split}
\end{align}
where the subscripts $p$ and $f$ stand for perturbed and fixed bunch, respectively. It can be seen that $\xi_{x,p}\sim\beta_{x,p}^*$ and $\xi_{x,f}\sim1/(\beta_{x,p}^*\varepsilon_{x,p})$. When increasing $\beta_{x,p}^*$, the horizontal beam-beam force experienced by the perturbed bunch increases, causing it to blow up more in comparison to when $\varepsilon_{x,p}$ is increased, since $\xi_{x,p}$ does not depend directly on $\varepsilon_{x,p}$. On the contrary, the fixed bunch behaves the same way when scanning $\beta_{x,p}^*$ or $\varepsilon_{x,p}$, since $\xi_{x,f}$ is inversely proportional to both parameters.

\begin{figure}[h]
	\centering\includegraphics[width=\columnwidth]{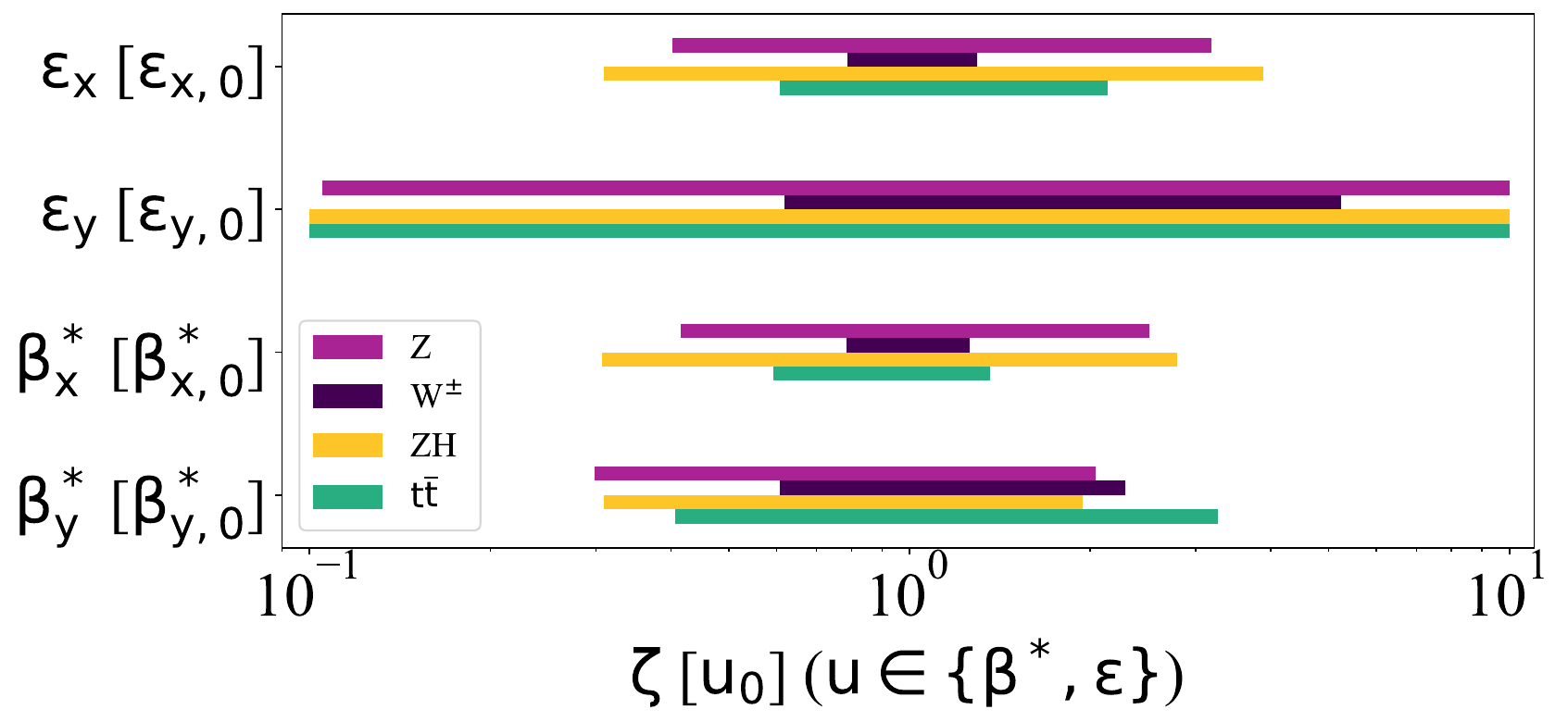}
    \caption{Visual overview of the limits on the asymmetry factor $\zeta$ (Eq.~\eqref{eq:uscan}) of various beam parameters in units of their nominal value from Tab.~\ref{tab:FCC}. The limits are defined where the vertical r.m.s. equilibrium of either of the bunches blows up by 50~\%.}
    \label{fig:blowups_50_percent}
\end{figure}

We define the tolerance limit of the beam parameters with respect to the flip-flop effect based on the equilibrium vertical r.m.s. blowing up by 50~\%, with respect to the equilibrium r.m.s. using the nominal parameters. These tolerance estimates can be useful for optics correction and tuning. The range for each parameter that can be considered safe by this definition is shown in Fig.~\ref{fig:blowups_50_percent}.

It can be seen that the blowup is the least sensitive to variations of $\varepsilon_y$ and in general more sensitive to variations in $\beta^*$ than in the emittance. As it was discussed earlier, the W$^{\pm}$ operation mode seems to be the most sensitive to parameter perturbations of any kind. The range of our parameter scans spans from 0.1 and 10 in all cases. Where the plotted bar extends all across this range indicates that the 50~\% blowup in $\sigma_y$ either never occurs or it takes place outside of this regime.

\section{\label{sec:6}Summary}

In this paper, we investigated the flip-flop effect in the context of the FCC-ee, using the \texttt{Xsuite} framework. We compared simulated equilibrium beam sizes with predictions from a theoretical model, which showed a remarkably good agreement for the equilibrium bunch length for realistically small bunch current asymmetries. In some cases we found a threshold in the bunch asymmetry with the onset of significant transverse blowup.

We have performed a first study of the currently proposed longitudinal top-up injection of the FCC-ee collider, using a linear lattice model. We have performed a series of parameter scans to assess the sensitivity of the dynamics to perturbations of several parameters, such as bunch intensity, emittance, optical beta function and injection offset. Our results show that in all FCC-ee operation modes the flip-flop mechanism does not have a detrimental effect on the collision luminosity with up to 5~\% initial asymmetry in the bunch currents (Fig.~\ref{fig:lumi}). The most sensitive to perturbations is the FCC-ee W$^{\pm}$ configuration. The sensitivity could likely be improved with an optimization of the configuration, such as a change of working point, yet these considerations go beyond the scope of this paper. We demonstrated that the injection offset can be safely increased up to the momentum acceptance and that this parameter has no significant effect on the luminosity or equilibrium beam sizes (Figs.~\ref{fig:topup_w} and~\ref{fig:topup_w_ev}). Our study with the bunch intensity asymmetry shows that the low energy operation modes Z and more importantly W$^{\pm}$ are more sensitive to the flip-flop mechanism (Figs.~\ref{fig:asy} and~\ref{fig:topup2}) than the higher energy setups. Furthermore, we have set up first estimates on the tolerance of beam emittance and optical $\beta$ function at the IPs, with respect to the flip-flop effect (Figs.~\ref{fig:beamsize_ex}, ~\ref{fig:beamsize_bx}, ~\ref{fig:beamsize_ey} and \ref{fig:beamsize_by}). 

In conclusion, our studies indicate that the interplay of the top-up injection and beam-beam collisions pose no show stoppers to the design of the FCC-ee. The next steps in this direction could include a more detailed investigation of the different trends in the blowup, as a function of the $\beta$ function and the emittance, as well as simulations of the top-up and beam-beam using a full element by element lattice model.

\acknowledgments{
The authors would like to thank D. Shatilov for helpful discussions on the benchmarking of the \texttt{Xsuite} beam-beam model, and M. Hofer for ideas on the top-up injection simulations. This work was carried out under the auspices of and with support from the Swiss Accelerator Research and Technology (CHART) programme (\url{www.chart.ch}) and with funding from the EU Future Cirtcular Collider Innovation Study (FCCIS) project under the grant agreement ID 951754 (\url{https://cordis.europa.eu/project/id/951754}).
}

\appendix

\nocite{*}

\bibliography{apssamp}

\end{document}